\documentclass[11pt,english,preprint]{aastex}
\usepackage[T1]{fontenc}
\usepackage[latin9]{inputenc}
\usepackage{geometry}
\usepackage{array}
\usepackage{rotfloat}
\usepackage{multirow}
\usepackage{amssymb}
\usepackage{graphicx}

\makeatletter

\providecommand{\tabularnewline}{\\}
\newcommand{\lyxdot}{.}

\usepackage{amsfonts}\usepackage{epstopdf}
\usepackage{amsmath}
\usepackage{esint}

\makeatother

\slugcomment{To Appear in The Astrophysical Journal}

\usepackage{babel}
\begin{document}

\title{Determining X-Ray Source Intensity and Confidence Bounds in Crowded
Fields}

\author{F. A. Primini and V. L. Kashyap}

\affil{Smithsonian Astrophysical Observatory, 60 Garden Street, Cambridge,
MA 02138}

\email{fap@head.cfa.harvard.edu}
\begin{abstract}
We present a rigorous description of the general problem
of aperture photometry in high energy astrophysics photon-count images,
in which the statistical noise model is Poisson, not Gaussian. We
compute the full posterior probability density function for the expected
source intensity for various cases of interest, including the important
cases in which both source and background apertures contain contributions
from the source, and when multiple source apertures partially overlap.
A Bayesian approach offers the advantages that it allows one to  (a) 
include explicit prior information on source intensities, (b)  propagate
posterior distributions as priors for future observations, and (c)
use Poisson likelihoods, making the treatment valid in the low
counts regime. Elements of this approach have been implemented in
the Chandra Source Catalog. 
\end{abstract}


\section{Introduction}

A common problem in astronomy is the estimate of the intensity of
a celestial source, using digital image data that also include contaminating
contributions from sky background and nearby sources. In optical,
infrared, and ultraviolet images, there are typically sufficient photon
events per pixel that a Gaussian statistical noise model can be assumed,
and one may fit a model spatial profile, including telescope response
and any intrinsic source extent, to the observed event distribution
\citep[see e.g.][]{Stetson:1987}. In X-ray and $\gamma$-ray images
however, there are typically few events per pixel, even for long exposures.
Moreover, the telescope response or point spread function (PSF) may
vary significantly with photon energy and with location in the field-of-view.
Its size may range from approximately one image pixel near the optical
axis to several tens of pixels at large off-axis distances. In such
cases, model-fitting to the sparse photon data can become difficult,
or at least computationally expensive, and researchers often resort
to simpler aperture photometry techniques. These involve counting
photon events in a region, or aperture, centered on the nominal source
location, with background determined from event counts in near-by
source-free regions. Net counts are then multiplied by correction
factors to convert counts to flux for an assumed spectral model and
to correct for losses due to detector/telescope efficiency or apertures
whose sizes do not enclose the full PSF at the source location. The
resulting intensities or fluxes are typically simple algebraic functions
of the raw aperture counts, and their errors are often estimated by
using simple propagation of error techniques which assume a Gaussian
statistical noise model.

A number of authors have attacked the problem using Bayesian statistical
techniques, which can naturally incorporate a Poisson noise model.
\citet{Loredo:1992} first pointed out the advantages of such techniques
to determine x-ray intensities for isolated sources, and \citet{Kraft:1991}
used a Bayesian formalism to determine confidence bounds on x-ray
intensities. Recently, \citet{Laird:2009} considered the astronomically
interesting case in which the prior distribution for source intensity
is given by a $logN-logS$ distribution, and showed that this can
naturally account for the sampling bias in intensity near detection
threshold. However, these treatments all assume that background is
either negligible or known and that background apertures are uncontaminated
by source counts. \citet{Weisskopf:2007} carried out a likelihood-based
analysis that treats the case where both source and background apertures
contain source contributions, and allows for uncertainties in background
measurements. However, their analysis only treats the case of isolated
sources and does not consider any prior information on source or background
intensity.

In this paper we present a full Bayesian treatment for the problem
and explicitly account for contributions from multiple sources in
both source and background apertures. We emphasize that we are addressing
the problem of estimating the range in which a source intensity is
likely to be found, at some given probability level, not the probability
that the source is real. The latter is an equally important but separate
problem \citep{Kashyap:2010}. We begin in Section~\ref{classical}
with a discussion of the Maximum Likelihood solution to ground the
user in our terminology. In Section~\ref{bayes:single} we present
our Bayesian formalism for the case of an isolated source and extend
the treatment to multiple sources in Section~\ref{bayes:multiple}.
In Section \ref{sec:Test-Cases}, we consider some examples and explore
the range of situations where our treatment is useful, using simulations.
We present the detailed mathematics of our derivations in appendices.

\section{Maximum Likelihood Estimate for Net Counts}

\label{classical}

We derive here the relevant formulae for computing maximum likelihood
estimates for net counts for an unresolved source or sources from
quantities obtained in aperture photometry measurements. We limit
our discussion to net counts but note that other quantities such as
source rate or flux can also be accommodated by introducing the appropriate
conversion factors (e.g. exposure or effective area). This section
essentially paraphrases the results derived in Appendix A of \citet{Weisskopf:2007},
modified only to accomodate the different variables and terms that
we use throughout the paper. These are defined in Table~\ref{tab1}.

\subsection{An Isolated Source}

\label{classical-isolated} 
\begin{figure}
\centering

\includegraphics[clip,width=6.5in]{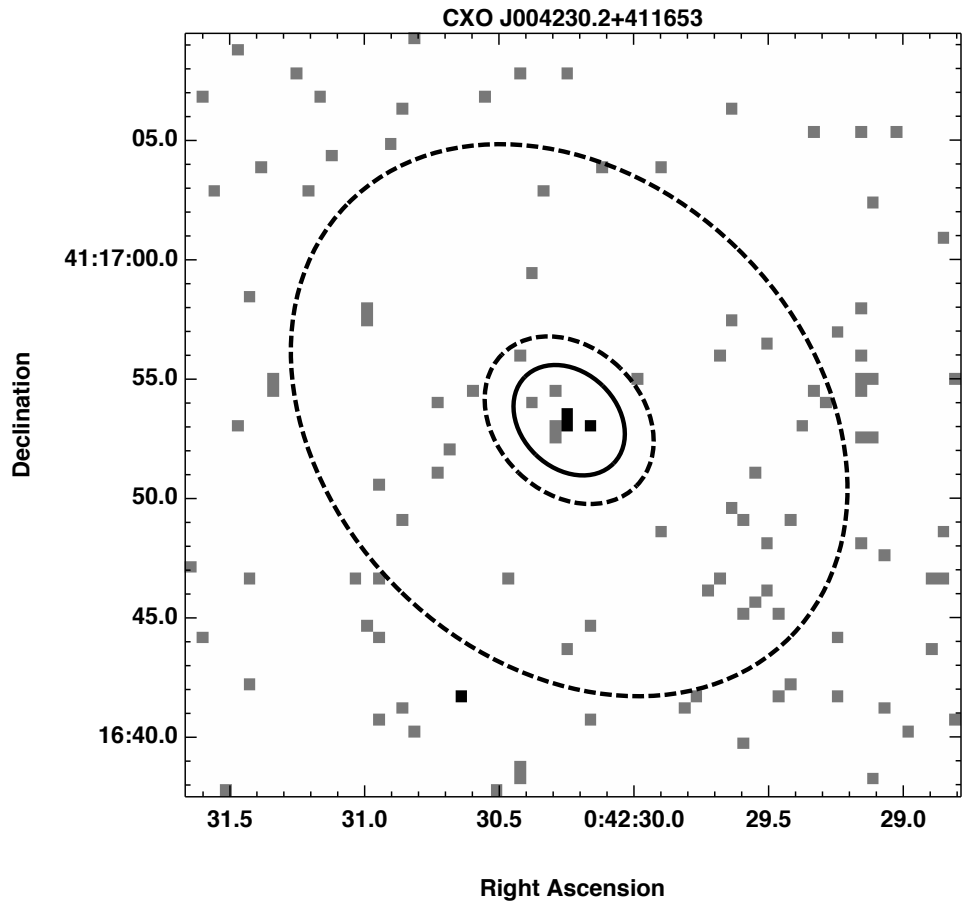}

\caption{\textit{Source (solid ellipse) and background apertures (dashed ellipses)
for an isolated X-ray source, from data obtained from Release 1.1
of the Chandra Source Catalog \citep{evans09}. The background aperture
has been modified slightly to illustrate the use of a detached aperture.
For this source, C=12, $\Omega_{s}$=67.74 $\, pix^{2}$, f=0.93,
B=33, $\Omega_{b}$=1537.41 $\, pix^{2}$, and g=0.03. \label{fig1}}}
\end{figure}

We consider first the simple case of a
single, isolated source, for which suitable source and background
apertures can be constructed without encountering other contaminating
sources. The situation is shown in Figure~\ref{fig1}. For clarity,
we omit the source subscript $i$. Although apertures may be of arbitrary
shape, subject to the limitation that $\int_{R}psf(X,Y,x,y)dxdy$
exist, we use apertures bounded by ellipses since they roughly approximate
the general shape of psfs for typical x-ray telescopes.

The ability to construct a suitable background aperture depends on
a balance of competing factors. In x-ray images with very low background
densities, it may be necessary to require $\Omega_{b}\gg\Omega_{s}$
in order to obtain an accurate measure of the background. One may
also wish to separate or detach the source and background apertures,
as we show in Figure~\ref{fig1}, to minimize the source contribution
to the background aperture. However, spatial variations in the background
and a high source density may force a smaller background aperture
situated close to the source, in order to approximate the background
with a constant value and to treat the source as isolated.

Assuming that appropriate apertures can be defined, the observed counts
$C$ in the source aperture and $B$ in the background aperture may
be treated as samples from Poisson distributions with means $\mu_{s}=fs+\Omega_{s}b$
and $\mu_{b}=gs+\Omega_{b}b$, where $f$ and $g$ are PSF fractions
in source and background apertures with areas $\Omega_{s}$ and $\Omega_{b}$,
and $s$ and $b$ are true source counts and background density, respectively.%
\footnote{We assume, for simplicity, that the exposures $E_{s}$ and $E_{b}$
in the source and background apertures are the same. This assumption
may be lifted by defining $s$ and $b$ as source rate and background
rate per unit area, and replacing $s$ and $b$ by the quantities
$s\times E_{s}$ and $b\times E_{b}$ . They can be similarly generalized
for source and background fluxes for given effective areas $\epsilon_{s}$
and $\epsilon_{b}$.%
} Since $C$ and $B$ are statistically independent, the total probability
of obtaining $C$ counts in source aperture $R_{s}$ and $B$ counts
in background aperture $R_{b}$ is given by

\begin{equation}
P(C,B|\mu_{s},\mu_{b})=Pois(\, C\,|\,\mu_{s})\, Pois(\, B\,|\,\mu_{b})=\frac{\mu_{s}^{C}}{\Gamma(C+1)}e^{-\mu_{s}}\frac{\mu_{b}^{B}}{\Gamma(B+1)}e^{-\mu_{b}}.\label{eq1}
\end{equation}
\\
Defining the log-likelihood function $L$ as

\begin{equation}
L=ln[P(C,B|\mu_{s},\mu_{b})]=C\, ln(\mu_{s})-\mu_{s}+B\, ln(\mu_{b})-\mu_{b}-ln[\Gamma(C+1)]-ln[\Gamma(B+1)],\label{eq:lnL-onesource}
\end{equation}
we obtain maximum-likelihood estimators for $s$ and $b$ by requiring
$\frac{\partial L}{\partial s}=0$ and $\frac{\partial L}{\partial b}=0$
. Both conditions are are satisfied by the solution to the two simultaneous
linear equations
\begin{eqnarray}
C & = & \mu_{s}=fs+\Omega_{s}b\nonumber \\
\label{eq:mu-single}\\
B & = & \mu_{b}=gs+\Omega_{b}b.\nonumber 
\end{eqnarray}
\\
The maximum-likelihood estimators for $s$ and $b$ (cf. \citet{Weisskopf:2007},
eq. A12 \& A13) are thus

\begin{eqnarray}
\hat{s} & = & \frac{C\Omega_{b}-B\Omega_{s}}{f\Omega_{b}-g\Omega_{s}}\nonumber \\
\label{eq:ML-single}\\
\hat{b} & = & \frac{Bf-Cg}{f\Omega_{b}-g\Omega_{s}}.\nonumber 
\end{eqnarray}
\\
When $C$ and $B$ are large, so that we can assume a Gaussian statistical
model, we can estimate the error in $\hat{s}$ and $\hat{b}$ using
simple propagation of errors:
\begin{eqnarray}
\sigma_{\hat{s}}^{2} & = & \frac{C\Omega_{b}^{2}+B\Omega_{s}^{2}}{(f\Omega_{b}-g\Omega_{s})^{2}}\nonumber \\
\label{eq:ML-single-sigma}\\
\sigma_{\hat{b}}^{2} & = & \frac{Bf^{2}+Cg^{2}}{(f\Omega_{b}-g\Omega_{s})^{2}}.\nonumber 
\end{eqnarray}

\subsection{Multiple Sources}

\label{classical-multiple} 
\begin{figure}[h!]
\centering

\includegraphics{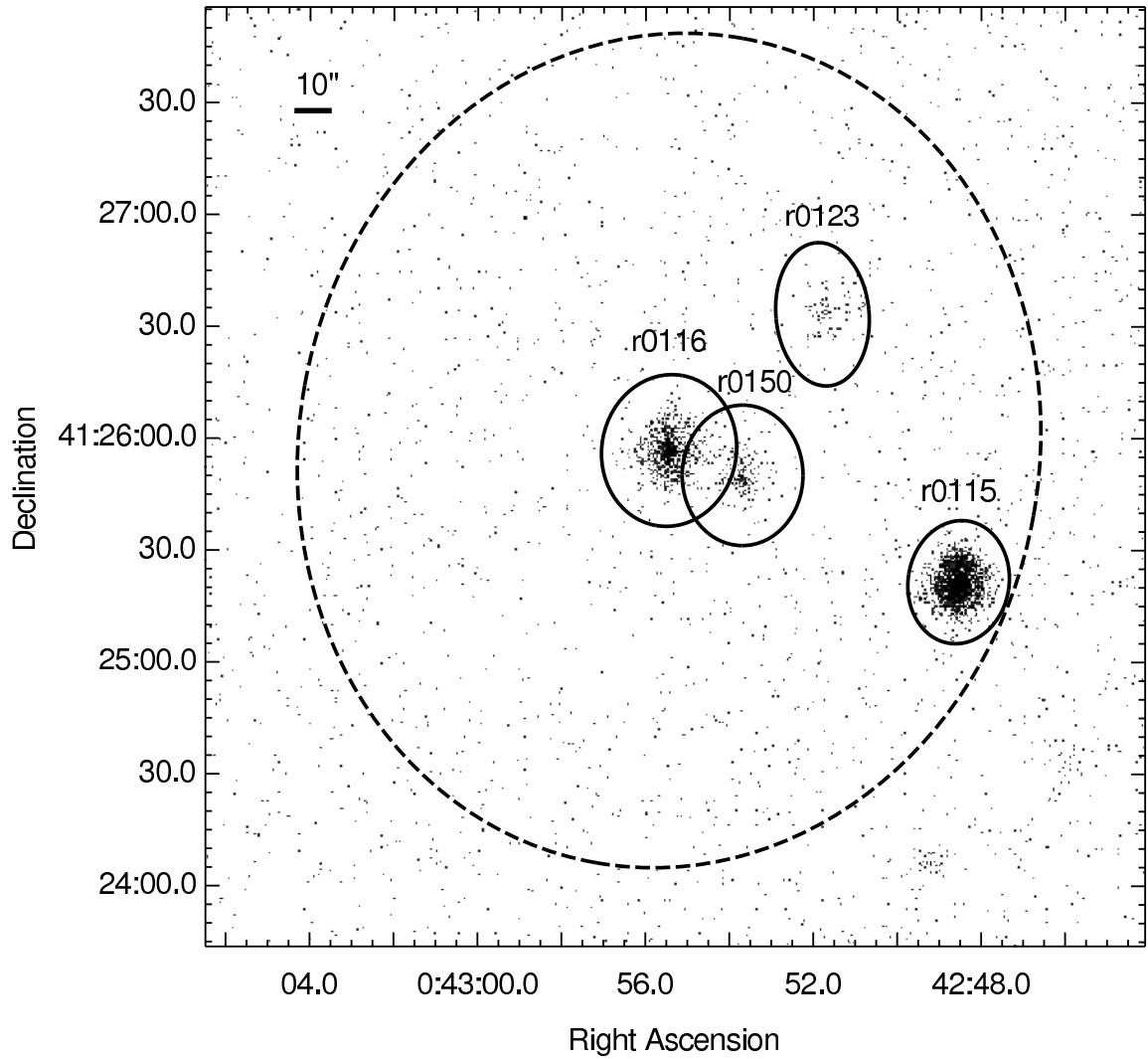} \caption{\textit{Source (solid ellipses) and background (dashed ellipse) apertures
for four sources in a crowded region of Chandra OBSID 1575, from data
obtained from Release 1.1 of the Chandra Source Catalog \citep{evans09}.
Source aperture labels correspond to the Region IDs described in Table
\ref{tab:region-data}. Data within the source apertures are excluded
from the background aperture.\label{fig:four_source_fig}}}
\end{figure}

Next, we consider the case in which there
are two or more sources which contribute to the counts in the source
and background apertures. The situation is illustrated in Figure~\ref{fig:four_source_fig}.
If the source apertures overlap, as is the case for two of the sources
here, events in the overlap region should be attributed to only one
of the overlapping source apertures, to preserve the statistical independence
of the aperture counts.\footnote{
An alternative approach for dealing with overlapping
apertures is suggested by \citet{2010ApJ...714.1582B} for the {\em ACIS 
Extract} software package. In that package, the aperture of the brighter
source remains unchanged, while that of the fainter source is repeatedly reduced in
size to include ever-decreasing encircled energy fractions, until the
overlap is eliminated. We discuss this approach further in Section~\ref{sec:newprocedure}.}
 Then, for $n$ sources, the log-likelihood
function $L$ is a simple extension to Equation~\ref{eq:lnL-onesource}:
\begin{equation}
L=\sum_{i=1}^{n}\left\{ C_{i}\, ln(\mu_{s_{i}})-\mu_{s_{i}}-ln[\Gamma(C_{i}+1)]\right\} +B\, ln(\mu_{b})-\mu_{b}-ln[\Gamma(B+1)],
\end{equation}
\\
and the maximum-likelihood estimators for $s_{i}$ and $b$ are obtained
by requiring that $\frac{\partial L}{\partial s_{i}}=0$ and $\frac{\partial L}{\partial b}=0$
. These conditions are satisfied by the solution to the set of $n+1$ simultaneous
linear equations (cf. \citealp{1994AIPC..313..239K})
\begin{eqnarray}
C_{i} & = & \mu_{s_{i}}  =  \sum_{j=1}^{n}f_{ij}s_{j}+\Omega_{s_{i}}b\nonumber \\
\label{eq:ML-mult}\\
B & = & \mu_{b}  =  \sum_{i=1}^{n}g_{i}s_{i}+\Omega_{b}b.\nonumber 
\end{eqnarray}
\\
Equation \ref{eq:ML-mult} can be written in matrix form as $\overline{C}=\overline{\overline{F}}\times\overline{S,}$
where vectors $\overline{C}$ and $\overline{S}$ are given by
\begin{eqnarray*}
\overline{C} & = & (C_{1},\ldots C_{n},\, B)\\
\overline{S} & = & (s_{1},\ldots s_{n},\, b),
\end{eqnarray*}
\\
and the matrix $\overline{\overline{F}}$ is given by$ $
\begin{eqnarray*}
\overline{\overline{F}} & = & \begin{bmatrix}f_{11} & \cdots & f_{1n} & \Omega_{s_{1}}\\
\vdots & \ddots & \vdots & \vdots\\
f_{n1} & \cdots & f_{nn} & \Omega_{s_{n}}\\
g_{1} & \cdots & g_{n} & \Omega_{b}
\end{bmatrix}.
\end{eqnarray*}
\\
The solution is then $\overline{S}=\overline{\overline{F^{-1}}}\times\overline{C}$
, where $\overline{\overline{F^{-1}}}$ is the inverse of $\overline{\overline{F}}$,
or
\begin{eqnarray}
\hat{s}_{k} & = & \sum_{j=1}^{n}F_{kj}^{-1}C_{j}+F_{k,n+1}^{-1}B\nonumber \\
\hat{b} & = & \sum_{j=1}^{n}F_{n+1,j}^{-1}C_{j}+F_{n+1,n+1}^{-1}B,\label{eq:ML-mult-matrix}
\end{eqnarray}
and the uncertainties are given by 
\begin{eqnarray}
\sigma_{\hat{s}_{k}}^{2} & = & \sum_{j=1}^{n}(F_{kj}^{-1})^{2}C_{j}+(F_{k,n+1}^{-1})^{2}B\nonumber \\
\sigma_{\hat{b}}^{2} & = & \sum_{j=1}^{n}(F_{n+1,j}^{-1})^{2}C_{j}+(F_{n+1,n+1}^{-1})^{2}B.\label{eq:ML-mult-sigma-matrix}
\end{eqnarray}

\section{Bayesian Formalism}

We now consider the problem from a Bayesian perspective. Our goal
is to derive relations for the posterior probability distributions
for background and source intensities which can be used to determine
intensities and credible regions analagous to the quantities described
in Equations \ref{eq:ML-single},\ref{eq:ML-single-sigma},\ref{eq:ML-mult-matrix},
and \ref{eq:ML-mult-sigma-matrix}.

\subsection{An Isolated Source}

\label{bayes:single}

We consider again the situation shown in Figure \ref{eq1}. We still
assume that the counts in the source and background apertures are
drawn from independent Poisson processes, but now use Bayes' Theorem
to express the posterior probability distributions for $\mu_{s}$and
$\mu_{b}$, the total intensities due to both source and background
in the respective apertures: 
\begin{eqnarray}
P(\mu_{s},\,\mu_{b}|\, C,\, B) & = & \frac{P(\mu_{s},\,\mu_{b})\, P(C,\, B\,|\,\mu_{s},\,\mu_{b})}{P(C,\, B)}\nonumber \\
\label{bayestheorem}\\
 & = & \frac{P(\mu_{s})P(\mu_{b})}{P(C,\, B)}\,\frac{\mu_{s}^{C}}{\Gamma(C+1)}e^{-\mu_{s}}\frac{\mu_{b}^{B}}{\Gamma(B+1)}e^{-\mu_{b}},\nonumber 
\end{eqnarray}
\\
where we have used the Poisson Likelihoods from Equation \ref{eq1}
and have taken advantage of the statistical independence of $C\,,B$
and $\mu_{s}\,,\mu_{b}$. For the prior probabilities for $\mu_{s}$
and $\mu_{b}$ we use $\gamma$-distributions of the form 
\begin{eqnarray}
P(\mu_{s}) & = & \frac{\beta_{s}^{\alpha_{s}}\mu_{s}^{\alpha_{s}-1}e^{-\beta_{s}\mu_{s}}}{\Gamma(\alpha_{s})},\nonumber \\
\label{eq:gamma-dist}\\
P(\mu_{b}) & = & \frac{\beta_{b}^{\alpha_{b}}\mu_{b}^{\alpha_{b}-1}e^{-\beta_{b}\mu_{b}}}{\Gamma(\alpha_{b})}.\nonumber 
\end{eqnarray}
\\
These distributions are referred to as conjugate priors for Poisson
Likelihood functions, since they result in posterior distributions
of the same functional form \citep{raiffa1961}. They are highly flexible
functions that can be used to specify the Poisson intensity a priori.
The number of counts is specified as $ $$\alpha-1$, and the relative
areas and exposure times are specified via $\beta$. In the limit
in which $\alpha_{s}\,,\alpha_{b}\rightarrow1$ and $\beta_{s}\,,\beta_{b}\rightarrow0$,
these approach non-informative, flat priors.

\begin{figure}
\centering

\includegraphics{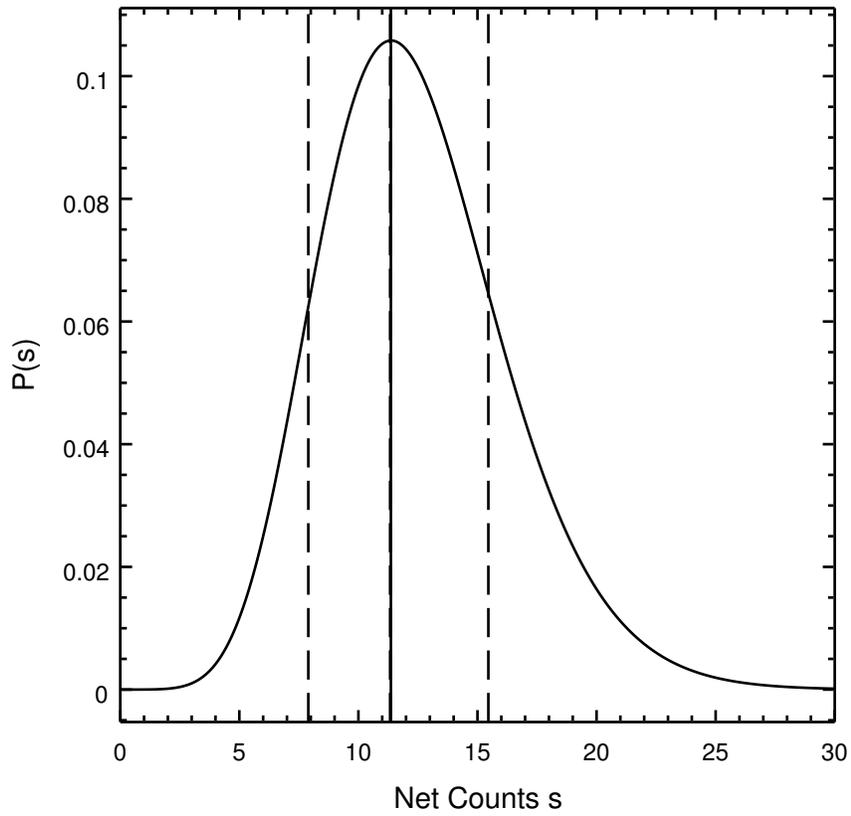}

\caption{\textit{Posterior probability distribution for the source shown in
Figure \ref{fig1}, evaluated using Equation \ref{eq:15}. The distribution
mode and 68\% confidence bounds are indicated with vertical dashed
lines. The Maximum Likelihood estimate is indicated by a solid vertical
line.\label{fig:Posterior-single}}}

\end{figure}

The joint posterior probability distribution is then
\begin{eqnarray}
P(\mu_{s},\,\mu_{b}|\, C,\, B\,) & = & \mu_{s}^{C+\alpha_{s}-1}e^{-\mu_{s}(1+\beta_{s})}\mu_{b}^{B+\alpha_{b}-1}e^{-\mu_{b}(1+\beta_{b})}\times\nonumber \\
\label{jpost-single}\\
 &  & \frac{1}{P(C,\, B)}\frac{\beta_{s}^{\alpha_{s}}\beta_{b}^{\alpha_{b}}}{\Gamma(\alpha_{s})\Gamma(C+1)\Gamma(\alpha_{b})\Gamma(B+1)}.\nonumber 
\end{eqnarray}
\\
The evidence term $P(C,B)$ is determined by the standard normalization
requirement
\begin{equation}
\intop_{0}^{\infty}d\mu_{s}\intop_{0}^{\infty}d\mu_{b}P(\mu_{s},\,\mu_{b}|\, C,\, B\,)=1,
\end{equation}
\\
and the posterior distribution $P(s)$ is determined by changing variables
from $\mu_{s}\,,\mu_{b}$ to $s\,,b$ and then marginalizing over
all values of b:
\begin{equation}
P(s\,|C,B)=\intop_{0}^{\infty}db\, P(s,b\,|C,B).
\end{equation}
 The mathematical details are provided in Appendix \ref{sec:Derivation-of-Posterior}.
The final result is
\begin{eqnarray}
P(s\,|C,\, B) & = & (\Omega_{b}f-\Omega_{s}g)\times(1+\beta_{s})^{(C+\alpha_{s})}\, e^{-fs(1+\beta_{s})}\times(1+\beta_{b})^{(B+\alpha_{b})}\, e^{-gs(1+\beta_{b})}\times\nonumber \\
\nonumber \\
 &  & \sum_{k=0}^{(C+\alpha_{s}-1)}\,\sum_{j=0}^{(B+\alpha_{b}-1)}\frac{(fs)^{k}\,\Omega_{s}^{(C+\alpha_{s}-1-k)}}{\Gamma(k+1)\Gamma(C+\alpha_{s}-k)}\times\frac{(gs)^{j}\,\Omega_{b}^{(B+\alpha_{b}-1-j)}}{\Gamma(j+1)\Gamma(B+\alpha_{b}-j)}\times\\
\nonumber \\
 &  & \frac{\Gamma(C+\alpha_{s}-k+B+\alpha_{b}-j-1)}{[\Omega_{s}(1+\beta_{s})+\Omega_{b}(1+\beta_{b})]^{(C+\alpha_{s}-k+B+\alpha_{b}-j-1)}}\,.\nonumber 
\end{eqnarray}
\\
For the case of non-informative prior distributions, with $\alpha_{s}=\alpha_{b}=1$
and $\beta_{s}=\beta_{b}=0$,
\begin{eqnarray}
P(s\,|C,\, B) & = & (\Omega_{b}f-\Omega_{s}g)\times\, e^{-fs}\times e^{-gs}\times\nonumber \\
\nonumber \\
 &  & \sum_{k=0}^{C}\,\sum_{j=0}^{B}\frac{(fs)^{k}\,\Omega_{s}^{(C-k)}}{\Gamma(k+1)\Gamma(C-k+1)}\times\frac{(gs)^{j}\,\Omega_{b}^{(B-j)}}{\Gamma(j+1)\Gamma(B-j+1)}\times\label{eq:15}\\
\nonumber \\
 &  & \frac{\Gamma(C-k+B-j+1)}{(\Omega_{s}+\Omega_{b})^{(C-k+B-j+1)}}\,.\nonumber 
\end{eqnarray}
\\
We use Equation~\ref{eq:15} to evaluate the posterior distribution for the
source shown in Figure \ref{fig1}. The result is shown in
Figure~\ref{fig:Posterior-single}. We note that the mode of the posterior distribution
is indistinguishable from the maximum likelihood estimate for net source
counts, as should be expected,  
since we assumed non-informative or flat priors in deriving
Equation~\ref{eq:15}. In such cases, as can be seen from
Equation~\ref{bayestheorem}, the posterior probability distribution reduces to
the product of the likelihoods. We shall examine this topic in more detail in Section~\ref{sec:Test-Cases}.

\subsection{Multiple Sources}

\label{bayes:multiple}We now consider multiple sources from a Bayesian
perspective. As before, Bayes' Theorem is used to express the joint
posterior probability distribution in terms of likelihoods and prior
probabilities. The details are provided in Appendix \ref{sec:Appendix 2}.
The marginalized posterior probability distribution for source $s_{i}$
is given in Equation \ref{eq:a2_final} as 
\begin{equation}
P(s_{i}\,|\, C_{1}\ldots C_{n}\,,\, B)ds_{i}=K^{\prime}\dotsintop_{b\,,\, s_{j}\neq s_{i}}db\, P(\mu_{b})\, Pois(B\,|\,\mu_{b})\prod_{i=1}^{n}ds_{i}P(\mu_{s_{i}})\, Pois(C_{i}\,|\,\mu_{s_{i}}).\label{eq:3.2_1}
\end{equation}
\\
A similar result holds for $P(b\,|\, C_{1}\ldots C_{n}\,,\, B)db$,
where integration is now over all sources, but not background.

We again assume $\gamma$-distributions for priors, so that, e.g.,
\begin{equation}
P(\mu_{s_{i}})=\frac{\beta_{i}^{\alpha_{i}}\mu_{s_{i}}^{\alpha_{i}-1}e^{-\beta_{i}\mu_{s_{i}}}}{\Gamma(\alpha_{i})}.\label{eq:3}
\end{equation}
\\
Since binomial expansions of powers containing
$\alpha_{i}$ are no longer used in evaluating marginalizing integrals
(as in Appendix \ref{sec:Derivation-of-Posterior}, Equation \ref{eq:binomial}),
the restriction that $\alpha_{i}$ and $\alpha_{b}$ be integers is
lifted.

The multiplicative constants in the prior distributions can be
absorbed into the single normalization constant $K^{\prime},$ yielding
\begin{equation}
P(s_{i}\,|\, C_{1}\ldots C_{n}\,,\, B)ds_{i}=K^{\prime}\dotsintop_{b\,,\, s_{j}\neq s_{i}}db\,\mu_{b}^{\alpha_{b}-1}e^{-\beta_{b}\mu_{b}}\, Pois(B\,|\,\mu_{b})\prod_{i=1}^{n}ds_{i}\mu_{s_{i}}^{\alpha_{i}-1}e^{-\beta_{i}\mu_{s_{i}}}\, Pois(C_{i}\,|\,\mu_{s_{i}}).\label{eq:4}
\end{equation}

As seen in Figure \ref{fig:Posterior-single}, the posterior distributions
are expected to be localized near the distribution mode, and to vary
smoothly. In such cases, it may be possible to evaluate the integrand
in Equation \ref{eq:4} on a suitable (n+1)-dimensional grid and evaluate
the n-dimensional marginalization integral by repeated one-dimensional
numerical integrations. In our web page\footnote{ \emph{hea-www.harvard.edu/XAP}},
we present  a sample Python program for doing
just that, using the maximum-likelihood estimates of source counts and errors to
define the parameters of the mesh. In the next section we use our code to
explore a number of test cases.

\section{Verification and Simulations\label{sec:Test-Cases}}

\subsection{Exemplar Test Cases}

In this section, we apply the procedure discussed at the end of the last
section to two test cases, using data from real {\em Chandra}
observations.

\subsubsection{An Isolated Point Source}

We begin with the simple case shown in Figure \ref{fig1}. As described at the
end of Section~\ref{bayes:single}, we computed
$P(s\,|\, C,\, B)$ analytically for the aperture data given in the
caption to Figure~\ref{fig1}, using Equation \ref{eq:15}, as implemented
in the \emph{CIAO} tool \emph{aprates}. We now use our new sample code to compute $P(s\,|\, C,\, B)$
numerically from Equation \ref{eq:4}. In both cases, we assumed non-informative
$\gamma-$distribution priors with $\alpha=1$ and $\beta=0$.  We compare the posterior distributions in
Figure 4. The distributions are in excellent agreement, demonstrating that our
numerical integration procedure and sample code produce results consistent with the analytical
result in the simple case where both are applicable.

\subsubsection{Sources in a Crowded Region}
We next consider the four {\em Chandra} Source Catalog sources shown in Figure \ref{fig:four_source_fig}.
All sources are treated at once, although only two have overlapping
apertures. However, one of the remaining sources, r0115, is sufficiently
bright that it may influence the background data even if its source
aperture is excluded from the background. Source and background data
for this case are listed in Table~\ref{tab:region-data}. For the
sources with overlapping source apertures, we have attributed counts
and area in the overlap region to the fainter of the two sources,
r0150. 

\paragraph{Non-Informative Priors}

We first assume 
{\em
non-informative}~\footnote{
Strictly speaking, there are no truly non-informative priors. Our
  choice of $\alpha_{i}=1$ and $\beta_{i}=0$ results in a flat, improper
  function in linear space. In some cases, a flat function in log space may be
  desired, or a formal least-information prior derived using the Fischer
  Information matrix. The choice of the putative non-informative prior has
  significant consequences for coverage rates (i.e., the frequency with which
  confidence bounds enclose true values) at low counts (see
  \citet{2006ApJ...652..610P}; see also Figures~\ref{fig:S1full_frac_err}~\&~\ref{fig:mle-frac-err}).
}
$\gamma$-distribution priors for all sources and background, 
with $\alpha_{i}=1$ and $\beta_{i}=0$, so that we can compare our results with
those of Release 1.1 of the {\em Chandra}
Source Catalog \citep{evans09}.
Our procedure yields the posterior distributions
shown in Figure \ref{fig:posterior-four}. To estimate confidence bounds, we approximate the mode
of each distribution as the vertex of a quadratic function fit to the three
highest points in the distribution. We then numerically integrate the sample posterior
distribution above and below the mode until the $68\%$
confidence bounds are obtained. For the two isolated sources, r0115
and r0123, the modes and confidence bounds, (black dashed vertical
lines), are in good agreement with those from Release 1.1 of the {\em Chandra}
Source Catalog (red dashed vertical lines), in which all sources were
treated independently. Results for the overlapping sources r0116 and
r0150 differ, as expected, since data in the overlap area were excluded
from the analysis in Release 1.1. At present, we only note that different
results are obtained. In Section~\ref{sec:Sims}, we present results of
simulations that demonstrate that the new procedure produces more accurate
results than that used in Release 1.1. 

\paragraph{Informative Priors}
We examine the effect of using informative priors by dividing the time
interval of the original data set
into two halves, and using the posterior distributions from one half (computed
assuming non-informative priors) to estimate the prior distributions for the
second. To do this, we note that, from the definition of
$\gamma$-distribution priors in Equation~\ref{eq:gamma-dist}
\begin{eqnarray}
\alpha&=&\frac{[E(\mu)]^2}{Var(\mu)}\nonumber \\
 \label{eq:def_alpha}\\
\beta&=&\frac{E(\mu)}{Var(\mu)}\nonumber
\end{eqnarray}
where 
\begin{eqnarray}
E(\mu)&=&\int_{0}^{\infty}d\mu\,\mu\,P(\mu)\nonumber \\
\label{eq:def_E_Var}\\
Var(\mu)&=&E(\mu^2)-[E(\mu)]^2\nonumber.
\end{eqnarray}
Since the aperture quantities $\mu_{s_{i}},\,\mu_b$ are linear combinations of
source and background intensities, as given in
Equation~\ref{eq:ML-mult} and Table~\ref{tab1}, we may write
\begin{eqnarray}
E(\mu_{s_{i}}) & = & \sum_{j=1}^{n}f_{ij}E(s_{j})+\Omega_{s_{i}}E(b)\nonumber \\
\label{eq:def_E}\\
Var(\mu_{s_{i}}) & = & \sum_{j=1}^{n}f_{ij}^{2}Var(s_{j})+\Omega_{s_{i}}^{2}Var(b)\,,\nonumber
\end{eqnarray}
and similarly for $E(\mu_b)$ and $Var(\mu_b)$.

We thus compute $E(s_i),\,Var(s_i),\,E(b),\,$ and $Var(b)$ from
Equation~\ref{eq:def_E_Var}, using the marginalized posterior distributions
$P(s_{i}\,|\, C_{1}\ldots C_{n}\,,\, B)$ and
$P(b\,|\,~C_{1}\ldots~C_{n}\,,\,B)$  from the first half of the data set as the probability distributions,
and use these to compute $E(\mu_{s_i}),\,Var(\mu_{s_i}),\,E(\mu_b),\,$ and
$Var(\mu_b)$ from Equation~\ref{eq:def_E}. These quantities are then used to
compute $\alpha_{s_{i}},\,\beta_{s_{i}},\,\alpha_{b},\,$ and $\beta_{b}$ from
Equation~\ref{eq:def_alpha} to define the prior distributions for analysis of
the second half of the data set.

Our results are shown in Figure~\ref{fig:inform-priors}. We note that for all four
sources the posterior distributions for the second half of the data set based
on informative priors are narrower than the equivalent distributions based on
non-informative priors, with modes consistent with the distributions derived
from the full dataset, based on non-informative priors. Note that by adopting
informative priors based on an analysis of the first half for the second half
of the observation, we make an implicit assumption that the sources do not
exhibit intrinsic variability; this assumption appears to be invalid for at
least one of the sources, r0116. 
 
Although it is tempting to err on the side of caution and include
all sources which may contribute to data in the background aperture,
there is a practical limit to the number of sources one can treat
at once in the simple numerical integration scheme that we use.
 The mesh size grows geometrically with the number of
sources, and must include an adequate number of points in any one
dimension to allow accurate determination of the mode and confidence
bounds. With a mesh size of $\sim20-30$ per source, current experience
indicates that fewer than 5 sources can be analyzed simultaneously
without exceeding typical memory resources. For example, analysis of 5 sources (a
6-dimensional mesh including background) with a mesh size of 30 per source
would require $\sim5$ Gbytes to hold  the joint posterior distribution in
memory. In such cases, more sophisticated
algorithms, such as Markov Chain Monte Carlo techniques, may be required
to evaluate Equation \ref{eq:4}. Alternatively, one may be able to ignore
sources in the joint computation based on their relative contributions. For
example, a source $j$ for which  $g_j\lesssim0.05$ and $f_{ij}\lesssim0.05$
for all other sources $i$ can likely be ignored since that is typically the
limit to which the point spread function is known.

\subsection{Limits of Applicability}
\label{sec:Sims}
Finally we investigate in more detail the performance of our procedure
using simulations. Our aim is to provide some comparison with other
techniques, and to explore the ranges in relative source
intensity and source separation for overlapping sources, for which our
procedure yields reliable results. 
\subsubsection{Simulation Set-Up}
We build a systematic grid for simulations based on source separation,
relative source intensity, and background level (D. Jones 2013, private communication).
We used the \textit{CIAO} tool \textit{ChaRT \citep{2003ASPC..295..477C}}\emph{,
Chandra }raytracing software \emph{SAOTrace \citep{2004SPIE.5165..402J},
}and \emph{CIAO} tools \emph{psf\_project\_ray} and \emph{dmcopy}
\citep{2006SPIE.6270E..60F} to generate an \emph{ACIS} image of the
point spread function for a source at an off-axis angle of $\sim0.5^{\prime}$
and pixel resolution of $\sim0.25^{\prime\prime}$, using the metadata
of Chandra observation 1575. We then used the two-dimensional modeling
capabilities of $Sherpa$ \citep{2001SPIE.4477...76F} to simulate
pairs of sources separated by $\Delta=0.5,1.0,1.5,2.0\times r_{90}$,
where $r_{90}$ is the average radius of an ellipse enclosing 90\%
of the encircled energy of the point spread function images, determined
using the \emph{CIAO} tool \emph{dmellipse}. At the image locations
chosen, $r_{90}\sim1^{\prime\prime}$. At each separation, we considered
a range of source intensities, with a bright source (source 1) with
model counts $M_{1}=1000$ and a fainter source (source 2) with model
counts $M_{2}=1000/r$. The relative intensity $r$ was chosen such
that $log_{10}(r)=0,0.5,1,1.5,2,$ corresponding to $M_{2}$ values
of $1000,\,316,\,100,\,31.6,$ and $10,$ respectively. Finally, we
considered three different background levels, with model background
in the 90\% encircled energy source aperture for source 2 set to $b\times900/r$,
with $b=0.001,\,0.010,\,0.100$. 
For each combination of $\Delta,\, r,$ and $b$ , we used $Sherpa$
to simulate 1000 images with appropriate statistics applied for background
and both source intensities. Examples for $r=1$ and $b=0.001$ are
shown in Figure~\ref{fig:sim04}. 

\subsubsection{Results for New Procedure}
\label{sec:newprocedure}
We analyzed each image with our
sample code, assuming non-informative priors for each source. We used
the 90\% encircled energy ellipses determined from $dmellipse$ to
define the source apertures, and a circular region centered between
the two sources with twenty-five times the area of a single source
aperture to define the background aperture. Such background aperture
sizes were typical of isolated point sources in Release 1.1 of the
CSC. For each combination of $\Delta,\, r,$ and $b$ , and for each
simulation $k$, we tabulated the modes, $S_{i}^{k}$, and 68\% confidence
bounds, $S_{i}^{k,-},S_{i}^{k,+}$, from the posterior probability
distributions for each source $i$ in the image, and computed the
average fractional error and fractional width, given by

\begin{eqnarray}
fractional\, error_{i} & = & \frac{1}{1000}\times\sum_{k=1}^{1000}(\, S_{i}^{k}-M_{i}\,)/M_{i}\nonumber\\
\label{eq:fracerr}\\
 fractional\, width_{i} & = & \frac{1}{1000}\times\sum_{k=1}^{1000}(S_{i}^{k,+}-S_{i}^{k,-})/M_{i}\nonumber
\end{eqnarray}
\\
where $M_{i}$ refers to $M_{1}$ and $M_{2}$ for source 1 and 2,
respectively.

For $\Delta\lesssim1.5\, r_{90}\,$, there is substantial overlap
in the source apertures, and we consider separately cases where overlap
area $\Omega_o$ is assigned to the aperture of source 1 (Case 1) and
source 2 (Case 2). To be specific, in Case 1 (for example), the aperture for
source 1 is the full 90\% encircled energy aperture with area $\Omega_{s_1}$,
which includes area $\Omega_o$. All counts that fall within $\Omega_{s_1}$ are
assigned to the aperture for source 1. Moreover, the aperture for source 2 is
reduced in area to be $\Omega_{s_2}-\Omega_o$, and only counts that fall
within this reduced area are assigned to the aperture for source 2. Case 2 is
defined similarly.   
Fractional errors for both
cases  are shown in Figure \ref{fig:S1full_frac_err}.
We display the results as sets of density plots and contour of fractional
error as a function of $\Delta$ and $log_{10}r$ for fixed values
of $b,$ using radial basis linear interpolation on a $4\times5$
~$\Delta-log_{10}r$ mesh to provide smooth images and contours. Since the
fractional errors, as defined in Equation~\ref{eq:fracerr}, could be negative,
we add a positive offset of 0.1 to all interpolated values to allow for a
logarithmic scaling in the density plots. Contour values are corrected for the
offset. Color
bars and contours are the same for all plots. To provide a basis for
comparison, we note that the intensity of an isolated point source
with negligible background has a statistical uncertainty of $\sim3\%$
for a $1000$ count source and $\sim10\%$ for a $100$ count source. 

As expected, fractional errors for source 1 are small over most of
the range of $\Delta$ and $log_{10}r$, exceeding $+5\%$ only for
$\Delta\lesssim0.75\, r_{90}$ and $log_{10}r\lesssim1$ (source 2
counts $\gtrsim100$). Fractional errors for the fainter source 2
are larger, and exceed $\sim+50\%$ for sources fainter than $\sim100$
counts or closer than $\sim r_{90}$ to source 1. It is interesting
to note that Case 1 yields better results for source 2 than Case 2 does.
For example, in Case 2 the fractional errors are in general larger in the region
$\Delta\lesssim1.0\, r_{90}$ and $log_{10}r\gtrsim1.5$ than in Case 1, and
the area in the density plots with fractional errors greater than $\sim+5\%$ is
larger in Case 2 than in Case 1. We attribute
this somewhat counter-intuitive efffect to the fact that the source
1 intensity, and hence its contribution to other aperture is more
accurately determined when overlap area (and hence all counts)  is assigned to its aperture.

Finally, in Figure \ref{fig:Average-fractional-width},
we show results for fractional widths of the posterior probability
distributions, displayed in a fashion similar to that used for fractional
errors, except that since the widths are positive-definite quantities, no
offset is added in displaying the density plots. For comparison, the $\pm1\sigma$ width for a $1000$ count
isolated point source with negligible background is $\sim6\%$. We
note again that better results for the fainter source 2 are achieved
for Case 1. For example, the fractional widths are in general smaller in the region
$\Delta\lesssim1.0\, r_{90}$ and $log_{10}r\gtrsim1.5$ than in Case 2, and
the area in the density plots with fractional widths greater than $\sim+50\%$ is
larger in Case 2 than in Case 1.

We emphasize that in our approach to resolving overlapping apertures, we do
not assign counts to particular sources, but rather to particular
apertures which have been modified to eliminate the overlap. Estimated counts
in all apertures, as indicated in Equation~\ref{eq:ML-mult} and
Table~\ref{tab1}, are modelled as a linear combination of background and all
source intensities, with proportionality constants determined by psf
contributions for sources and aperture area for background. Although it is
possible to treat the overlap area as an additional aperture, this
significantly complicates the mathematical treatment of the problem, and we
don't consider it here. \footnote{Since the
number of source apertures is then no longer the same as the number of
sources, the system of linear equations described in Equation~\ref{eq:ML-mult}
is over-determined, with no unique solution for maximum-likelihood extimators
for $s_i$ and $b$. Further, the Jacobean determinant used to change variables
in Equation~\ref{B2} is undefined since the Jacobean matrix is no longer square.}
We note that the major differences between Case 1 and Case 2, as indicated in
Figure~\ref{fig:S1full_frac_err}, occur for $\Delta<\sim1$. As a point of
reference, in Release 1 of the Chandra Source Catalog \citep{evans09}, these close pairs
amounted to fewer than $\sim1\%$ of the total number of sources on average, although the
fraction could be significantly larger in dense stellar clusters and nuclei of galaxies.

The approach of \citet{2010ApJ...714.1582B} is similar to our Case 1, in that
the aperture of the brighter source remains unchanged, while that of the
fainter source is reduced. However, differences
in the details of the reduced apertures may lead to
somewhat different results.

\subsubsection{Comparison with Maximum Likelihood Results}
We also computed the Maximum Likelihood values for source intensity and
uncertainty for both sources in each simulated image, using Equations~\ref{eq:ML-mult-matrix}
and \ref{eq:ML-mult-sigma-matrix}. We then computed average fractional errors and
widths as in Equation~\ref{eq:fracerr}, substituting $\hat{s}_{k} $ for $S^k$
and $2\times\sigma_{\hat{s}_{k}}$ for $(S^{k,+}-S^{k,-})$. Cases 1 and 2 were
defined as before. Our results are shown in Figures~\ref{fig:mle-frac-err}
and \ref{fig:mle-frac-width}, which may be compared to
Figures~\ref{fig:S1full_frac_err} and \ref{fig:Average-fractional-width},
respectively. The fractional errors for Source 1 and the fractional widths for
both sources are, in fact, comparable to those determined using our
procedure, for both Case 1 and Case 2. This might be expected, since we used
non-informative priors in our current analysis, and, as noted at the end of
Section~\ref{bayes:single}, in such cases the Bayesian formalism reduces to
the Maximum Likelihood one. However, for the fainter Source 2, the
Maximum Likelihood average fractional errors are, in fact, much lower than those
computed using our procedure in the region
$\Delta\lesssim1.5\, r_{90}$ and $log_{10}r\gtrsim1.0$ (Source 2 counts
$<100$). We attribute this to the fact that, although we use
``non-informative'' $\gamma-$distribution priors with $\alpha=1$ and
$\beta=0$, we do take advantage of some prior information in our procedure,
namely, the implicit assumption that all source intensities are
non-negative. For bright sources, this prior information is of little
significance, but for faint sources with few counts near brighter sources, it
could be. In contrast, Maximum Likelihood estimators for source intensity do
allow negative values, since they provide the most probable intensities for a
particular dataset. For faint sources, positive statistical fluctuations in background, combined
with negative statistical fluctuations in source counts, could lead to
negative source intensities in the absence of any prior
constraints. Indeed, in the region $\Delta\lesssim1.5\, r_{90}$ and
$log_{10}r\gtrsim1.0$, approximately half of the Maximum Likelihood solutions
for Source 2 intensity are negative. For those cases, the modes of the
posterior distributions determined from our procedure are 0. Since the
fractional errors defined in Equation~\ref{eq:fracerr} are signed quantities,
the averages for the Maximum Likelihood solutions will be less than those from
our procedure.
A similar effect was noted by \citet{2006ApJ...652..610P}, who find improved
results when using a $\gamma-$distribution prior that is flat in log space.

\subsubsection{Comparison with Chandra Source Catalog Release 1.1 Photometry}
Finally, we compare the results from our procedure with those expected from
the analysis procedure used in  Release 1.1 of the {\em Chandra}
Source Catalog \citep{evans09}. In that procedure, all sources are analyzed
individually, and near-by contaminating sources are accounted for by excluding
their entire source aperture from the background aperture and the aperture of
the source being analyzed. We can mimic that process in our procedure by
considering Source 1 and Source 2 separately, with appropriately chosen
apertures, namely $\Omega_{s_1}-\Omega_o$  (the Case 2 aperture for source 1)
when analyzing Source 1 and $\Omega_{s_2}-\Omega_o$ (the Case 1 aperture for
source 2) when analyzing
Source 2. The results are shown in Figure~\ref{fig:aprates}. Here,
the results for Source 1 in Figure~\ref{fig:aprates}a should be compared to
those for Source 1 in 
Figure~\ref{fig:S1full_frac_err}b and the results for Source 2 in Figure~\ref{fig:aprates}a should be compared with
those for Source 2 in Figure~\ref{fig:S1full_frac_err}a. The corresponding
comparisons for fractional width are Source 1 in  Figures~\ref{fig:aprates}b
and \ref{fig:Average-fractional-width}b, and Source 2 in Figures~\ref{fig:aprates}b
and \ref{fig:Average-fractional-width}a. In all cases, the fractional widths
are comparable in the two procedures, but fractional errors are smaller for
both sources using our current procedure.
 
\section{Summary}

We present a general Bayesian formalism for computing posterior distributions
of source intensity in crowded fields. Distributions of intensities
of multiple sources are determined simultaneously, through appropriate
marginalization integrals of the joint posterior probability distribution.
The procedure depends on the individual source point spread functions
only through their integral properties, and hence is likely to be
more robust than methods that depend on detailed psf fitting. We present
examples from real data and simulations to illustrate the performance
of the procedure and demonstrate that it duplicates the performance
of the current $CIAO\,\, aprates$ tool used in Release 1.1 of the
CSC for isolated sources. When source apertures overlap, the standard
calculation differs significantly from the posterior distributions calculated
by the new procedure. We carry out simulations to demonstrate the advantages
of the new procedure.

When  non-informative priors that are flat in linear space are used, our procedure yields
results comparable to a Maximum Likelihood analysis for brighter sources, although the latter method yields better results for fainter
  sources. Improved results may be obtained for our procedure through the use
  of non-informative priors that are flat in log space.

When informative priors are used, our procedure
can produce more accurate results. This may be particularly useful in
combining data from multiple observations, such as a mosaic, in which the
apertures and point spread functions for the same source may differ
significantly in the various observations. In such cases, in the absence of
variability, source intensity and uncertainty from one observation may be
used to define the prior distribution for a subsequent observation.  

In order to preserve statistical independence for all source apertures
(so that Equation \ref{eq:3.2_1} holds), the procedure requires that
areas in which two apertures overlap, and the counts contained in the overlap
area,  be assigned to only one aperture. Depending
on the number of sources involved, there may be many ways of assigning
overlap area. Results of our current simulations indicate that assigning
the overlap to the aperture of the brighter source is preferable, although this should
be verified with simulations of more complicated cases.

Finally, one must consider how many sources can be considered simultaneously.
As shown in the example in Figure \ref{fig:four_source_fig}, multiple
sources may be considered even when their source apertures do not
overlap. However, practical considerations may limit this number.
A simple numerical integration scheme, as we describe in Section \ref{sec:Test-Cases},
is suitable when the number of sources is few, but may severely tax
computer memory resources when the number is large. For such cases,
more sophisticated schemes, such as Markov Chain Monte Carlo techniques,
may be required.

\acknowledgements{}
We thank the anonymous referee for many useful comments and criticisms. We also acknowledge
useful discussions with Tom Loredo and members of the CHASC
AstroStatistics Collaboration, especially Alanna Connors, David van
Dyk, and David Jones.
Support for this work was provided by the Chandra X-ray Observatory Center,
which is operated by the Smithsonian Astrophysical Observatory for and on
behalf of the National Aeronautics Space Administration under contract
NAS8-03060. VLK also acknowledges support from Chandra grant AR0-11001X. 

\appendix{\pagebreak{}}

\section{Derivation of Posterior Probability Distribution for an Isolated
Source\label{sec:Derivation-of-Posterior}}

We determine the evidence term $P(C,\, B)$ by requiring $\intop_{0}^{\infty}d\mu_{s}\intop_{0}^{\infty}d\mu_{b}P(\mu_{s},\,\mu_{b}|\, C,\, B\,)=1.$
Since $\Gamma(A)=B^{A}\int_{0}^{\infty}dx\, x^{A-1}e^{-Bx},$ we find
\begin{eqnarray}
P(C,\, B) & = & \frac{\Gamma(C+\alpha_{s})\beta_{s}^{\alpha_{s}}\Gamma(B+\alpha_{b})\beta_{b}^{\alpha_{b}}}{\Gamma(\alpha_{s})\Gamma(C+1)(1+\beta_{s})^{(C+\alpha_{s})}\Gamma(\alpha_{b})\Gamma(B+1)(1+\beta_{b})^{(B+\alpha_{b})}},
\end{eqnarray}
\\
and

\begin{eqnarray}
P(\mu_{s},\,\mu_{b}|\, C,\, B\,) & = & \frac{(1+\beta_{s})^{(C+\alpha_{s})}\mu_{s}^{C+\alpha_{s}-1}e^{-\mu_{s}(1+\beta_{s})}}{\Gamma(C+\alpha_{s})}\times\frac{(1+\beta_{b})^{(B+\alpha_{b})}\mu_{b}^{B+\alpha_{b}-1}e^{-\mu_{b}(1+\beta_{b})}}{\Gamma(B+\alpha_{b})}.\label{eq:final-single-source-posterior}
\end{eqnarray}

In order to obtain the posterior probability distribution for source
intensity $s$, marginalized over all values of background intensity
$b,$ we integrate the joint posterior distribution over all values
of $b$, changing variables from $(\mu_{s},\,\mu_{b}),$ to $(s,\, b)$:
\begin{eqnarray}
\intop_{all\, b}d\mu_{s}d\mu_{b}P(\mu_{s},\,\mu_{b}|\, C,\, B\,) & = & \intop_{b=0}^{\infty}\frac{\partial(\mu_{s},\,\mu_{b})}{\partial(s,\, b)}\, ds\, db\, P(\mu_{s}(s,\, b),\,\mu_{b}(s,\, b)|\, C,\, B\,),\nonumber \\
\label{eq:marg1}\\
 & = & P(s\,|C,\, B)\, ds,\nonumber 
\end{eqnarray}
\\
where the Jacobian determinant is 
\begin{eqnarray}
\frac{\partial(\mu_{s},\,\mu_{b})}{\partial(s,\, b)} & = & \frac{\partial\mu_{s}}{\partial s}\frac{\partial\mu_{b}}{\partial b}-\frac{\partial\mu_{b}}{\partial s}\frac{\partial\mu_{s}}{\partial b}=\Omega_{b}f-\Omega_{s}g.\label{eq:jacobian-determinant}
\end{eqnarray}
\\
Thus, we have
\begin{eqnarray}
P(s\,|C,\, B)\, ds & = & ds\,(\Omega_{b}f-\Omega_{s}g)\frac{(1+\beta_{s})^{(C+\alpha_{s})}}{\Gamma(C+\alpha_{s})}\, e^{-fs(1+\beta_{s})}\times\frac{(1+\beta_{b})^{(B+\alpha_{b})}}{\Gamma(B+\alpha_{b})}\, e^{-gs(1+\beta_{b})}\times\nonumber \\
\nonumber \\
 &  & \intop_{0}^{\infty}db\,(fs+\Omega_{s}b)^{C+\alpha_{s}-1}e^{-\Omega_{s}b(1+\beta_{s})}(gs+\Omega_{b}b)^{B+\alpha_{b}-1}e^{-\Omega_{b}b(1+\beta_{b})}\nonumber \\
\label{psds}\\
 & = & ds\,(\Omega_{b}f-\Omega_{s}g)\frac{(1+\beta_{s})^{(C+\alpha_{s})}e^{-fs(1+\beta_{s})}}{\Gamma(C+\alpha_{s})}\times\frac{(1+\beta_{b})^{(B+\alpha_{b})}e^{-gs(1+\beta_{b})}}{\Gamma(B+\alpha_{b})}\times\nonumber \\
\nonumber \\
 &  & \intop_{0}^{\infty}db\,(fs+\Omega_{s}b)^{C+\alpha_{s}-1}(gs+\Omega_{b}b)^{B+\alpha_{b}-1}e^{-b[\Omega_{s}(1+\beta_{s})+\Omega_{b}(1+\beta_{b})]}.\nonumber 
\end{eqnarray}
\\
If we limit our choices for $\alpha_{s}$ and $\alpha_{b}$ to be
integers, we can use the Binomial Theorem to write
\begin{eqnarray}
(fs+\Omega_{s}b)^{C+\alpha_{s}-1} & = & \sum_{k=0}^{C+\alpha_{s}-1}\left(\begin{array}{c}
C+\alpha_{s}-1\\
k
\end{array}\right)\,(fs)^{k}\,(\Omega_{s}b)^{C+\alpha_{s}-1-k}\nonumber \\
\label{eq:binomial}\\
 & = & \sum_{k=0}^{C+\alpha_{s}-1}\frac{\Gamma(C+\alpha_{s})}{\Gamma(k+1)\Gamma(C+\alpha_{s}-k)}\,(fs)^{k}\,(\Omega_{s}b)^{C+\alpha_{s}-1-k}\nonumber 
\end{eqnarray}
$ $\\
and a similar expression for $(gs+\Omega_{b}b)^{B+\alpha_{b}-1}$.
Equation \ref{psds} can then be written 
\begin{eqnarray}
P(s\,|C,\, B)\, ds & = & ds\,(\Omega_{b}f-\Omega_{s}g)\times(1+\beta_{s})^{(C+\alpha_{s})}\, e^{-fs(1+\beta_{s})}\times(1+\beta_{b})^{(B+\alpha_{b})}\, e^{-gs(1+\beta_{b})}\times\nonumber \\
\nonumber \\
 &  & \sum_{k=0}^{(C+\alpha_{s}-1)}\,\sum_{j=0}^{(B+\alpha_{b}-1)}\frac{(fs)^{k}\,\Omega_{s}^{(C+\alpha_{s}-1-k)}}{\Gamma(k+1)\Gamma(C+\alpha_{s}-k)}\times\frac{(gs)^{j}\,\Omega_{b}^{(B+\alpha_{b}-1-j)}}{\Gamma(j+1)\Gamma(B+\alpha_{b}-j)}\times\\
\nonumber \\
 &  & \frac{\Gamma(C+\alpha_{s}-k+B+\alpha_{b}-j-1)}{[\Omega_{s}(1+\beta_{s})+\Omega_{b}(1+\beta_{b})]^{(C+\alpha_{s}-k+B+\alpha_{b}-j-1)}}\,.\nonumber 
\end{eqnarray}
\\
For the case of non-informative prior distributions, with $\alpha_{s}=\alpha_{b}=1$
and $\beta_{s}=\beta_{b}=0$, we have
\begin{eqnarray}
P(s\,|C,\, B)\, ds & = & ds\,(\Omega_{b}f-\Omega_{s}g)\times\, e^{-fs}\times e^{-gs}\times\nonumber \\
\nonumber \\
 &  & \sum_{k=0}^{C}\,\sum_{j=0}^{B}\frac{(fs)^{k}\,\Omega_{s}^{(C-k)}}{\Gamma(k+1)\Gamma(C-k+1)}\times\frac{(gs)^{j}\,\Omega_{b}^{(B-j)}}{\Gamma(j+1)\Gamma(B-j+1)}\times\nonumber \\
\nonumber \\
 &  & \frac{\Gamma(C-k+B-j+1)}{(\Omega_{s}+\Omega_{b})^{(C-k+B-j+1)}}\,,
\end{eqnarray}
\\
or 
\begin{eqnarray}
P(s\,|C,\, B)\, ds & = & ds\,(\Omega_{b}f-\Omega_{s}g)\times\nonumber \\
\label{eq:final_bayes_single}\\
 &  & \sum_{k=0}^{C}\,\sum_{j=0}^{B}\frac{Pois(k\,|\, fs)\,\Omega_{s}^{(C-k)}}{\Gamma(C-k+1)}\,\frac{Pois(j\,|\, gs)\,\Omega_{b}^{(B-j)}}{\Gamma(B-j+1)}\,\frac{\Gamma(C-k+B-j+1)}{(\Omega_{s}+\Omega_{b})^{(C-k+B-j+1)}}.\nonumber 
\end{eqnarray}

\section{Posterior Probability Distribution for Multiple Sources\label{sec:Appendix 2}}

Because of the additional mathematical complexity, we don't attempt
to derive an analytical expression for the joint posterior probability
distribution for $n$ sources plus background. Rather, we assume that
the marginalization integrals will be computed numerically, and take
advantage of a change in variables to evaluate the joint posterior
probability on an $(n+1)$-dimensional grid of $s_{1}\ldots s_{n},\, b$
, for easier marginalization.

We can extend Equation \ref{bayestheorem} to $n$ sources as 
\begin{eqnarray}
P(\mu_{s_{i}}\ldots\mu_{s_{n}},\,\mu_{b}|\, C_{1}\ldots C_{n},\, B) & = & K\, P(\mu_{b})\, Pois(B\,|\,\mu_{b})\times\nonumber \\
 &  & \prod_{i=1}^{n}P(\mu_{s_{i}})\, Pois(C_{i}\,|\,\mu_{s_{i}}),
\end{eqnarray}
\\
where the normalization constant $K$ includes the Evidence term.
We can then write the marginalization integral for source $s_{i}$
as
\begin{eqnarray}
P(s_{i}\,|\, C_{1}\ldots C_{n}\,,\, B)ds_{i} & = & K\dotsintop_{b\,,\, s_{j}\neq s_{i}}d\mu_{b}\, P(\mu_{b})\, Pois(B\,|\,\mu_{b})\prod_{i=1}^{n}d\mu_{s_{i}}P(\mu_{s_{i}})\, Pois(C_{i}\,|\,\mu_{s_{i}})\label{B2}\\
 & = & K\dotsintop_{b\,,\, s_{j}\neq s_{i}}\frac{\partial(\mu_{s_{1}}\ldots\mu_{s_{n}},\,\mu_{b})}{\partial(s_{1}\ldots s_{n},\, b)}\, db\, P(\mu_{b})\, Pois(B\,|\,\mu_{b})\prod_{i=1}^{n}ds_{i}P(\mu_{s_{i}})\, Pois(C_{i}\,|\,\mu_{s_{i}}).\nonumber 
\end{eqnarray}

We note that since $\mu_{s_{i}}$ and $\mu_{b}$ are linear functions
of $s_{1}\ldots s_{n}$ and $b$ (cf. Table \ref{tab1}), the $(n+1)$-dimensional
Jacobian determinant $\frac{\partial(\mu_{s_{1}}\ldots\mu_{s_{n}},\,\mu_{b})}{\partial(s_{1}\ldots s_{n},\, b)}$
is indepedent of $s_{1}\ldots s_{n}$ and $b$. For example, for the
case $n=2$, 
\begin{equation}
\frac{\partial(\mu_{s_{1}},\,\mu_{s_{2}},\,\mu_{b})}{\partial(s_{1},\, s_{2},\, b)}=f_{11}(f_{22}\Omega_{b}-g_{2}\Omega_{s_{2}})-f_{12}(f_{21}\Omega_{b}-g_{1}\Omega_{s_{2}})+\Omega_{s_{1}}(f_{21}g_{2}-f_{22}g_{1}).
\end{equation}
\\
It can therefore be absorbed into the normalization constant $K$,
and we can write
\begin{equation}
P(s_{i}\,|\, C_{1}\ldots C_{n}\,,\, B)ds_{i}=K^{\prime}\dotsintop_{b\,,\, s_{j}\neq s_{i}}db\, P(\mu_{b})\, Pois(B\,|\,\mu_{b})\prod_{i=1}^{n}ds_{i}P(\mu_{s_{i}})\, Pois(C_{i}\,|\,\mu_{s_{i}}).\label{eq:a2_final}
\end{equation}

\clearpage{}

\bibliographystyle{jwapjbib}
\nocite{*}
\bibliography{jw_abbrv,mnemonic,references,apj_abbrv}

\begin{thebibliography}{}

\bibitem[\protect\astroncite{{Broos} et~al.}{2010}]{2010ApJ...714.1582B}
{Broos}, P.~S., {Townsley}, L.~K., {Feigelson}, E.~D., {Getman}, K.~V.,
  {Bauer}, F.~E., \& {Garmire}, G.~P.,  2010, \apj, 714, 1582

\bibitem[\protect\astroncite{{Carter} et~al.}{2003}]{2003ASPC..295..477C}
{Carter}, C., {Karovska}, M., {Jerius}, D., {Glotfelty}, K., \& {Beikman}, S.,
  2003,
\newblock in Astronomical Data Analysis Software and Systems XII, ed.
  {H.~E.~Payne, R.~I.~Jedrzejewski, \& R.~N.~Hook}, Vol. 295,  477

\bibitem[\protect\astroncite{{Evans} et~al.}{2010}]{evans09}
{Evans}, I.~N., et~al., 2010, \apjs, 189, 37

\bibitem[\protect\astroncite{{Freeman}, {Doe} \&
  {Siemiginowska}}{2001}]{2001SPIE.4477...76F}
{Freeman}, P., {Doe}, S., \& {Siemiginowska}, A.,  2001,
\newblock in Society of Photo-Optical Instrumentation Engineers (SPIE)
  Conference Series, ed. J.-L. {Starck}, F.~D. {Murtagh}, Vol. 4477, 76

\bibitem[\protect\astroncite{{Fruscione} et~al.}{2006}]{2006SPIE.6270E..60F}
{Fruscione}, A., et~al., 2006,
\newblock in Society of Photo-Optical Instrumentation Engineers (SPIE)
  Conference Series, Vol. 6270

\bibitem[\protect\astroncite{{Jerius} et~al.}{2004}]{2004SPIE.5165..402J}
{Jerius}, D.~H., et~al., 2004,
\newblock in Society of Photo-Optical Instrumentation Engineers (SPIE)
  Conference Series, ed. K.~A. {Flanagan}, O.~H.~W. {Siegmund}, Vol. 5165, 402

\bibitem[\protect\astroncite{{Kashyap} et~al.}{1994}]{1994AIPC..313..239K}
{Kashyap}, V., {Micela}, G., {Sciortino}, S., {Harnden}, Jr., F.~R., \&
  {Rosner}, R.,  1994,
\newblock in The Soft X-ray Cosmos, ed. E.~M. {Schlegel}, R. {Petre}, Vol. 313,
   239

\bibitem[\protect\astroncite{{Kashyap} et~al.}{2010}]{Kashyap:2010}
{Kashyap}, V.~L., {van Dyk}, D.~A., {Connors}, A., {Freeman}, P.~E.,
  {Siemiginowska}, A., {Xu}, J., \& {Zezas}, A.,  2010, \apj, 719, 900

\bibitem[\protect\astroncite{{Kraft}, {Burrows} \& {Nousek}}{1991}]{Kraft:1991}
{Kraft}, R.~P., {Burrows}, D.~N., \& {Nousek}, J.~A.,  1991, \apj, 374, 344

\bibitem[\protect\astroncite{{Laird} et~al.}{2009}]{Laird:2009}
{Laird}, E.~S., et~al., 2009, \apjs, 180, 102

\bibitem[\protect\astroncite{{Loredo} \& {Wasserman}}{1993}]{Loredo:1993}
{Loredo}, T., \& {Wasserman}, I.,  1993,
\newblock in American Institute of Physics Conference Series, ed.
  {M.~Friedlander, N.~Gehrels, \& D.~J.~Macomb}, Vol. 280, 749

\bibitem[\protect\astroncite{{Loredo}}{1992}]{Loredo:1992}
{Loredo}, T.~J.,  1992,
\newblock in Statistical Challenges in Modern Astronomy, ed. {E.~D.~Feigelson
  \& G.~J.~Babu}, 275

\bibitem[\protect\astroncite{{Park} et~al.}{2006}]{2006ApJ...652..610P}
{Park}, T., {Kashyap}, V.~L., {Siemiginowska}, A., {van Dyk}, D.~A., {Zezas},
  A., {Heinke}, C., \& {Wargelin}, B.~J.,  2006, \apj, 652, 610

\bibitem[\protect\astroncite{Raiffa \& Schlaifer}{1961}]{raiffa1961}
Raiffa, H., \& Schlaifer, R.,  1961,
\newblock Applied Statistical Decision Theory,
\newblock  (Cambridge, Mass.: M.I.T. Press), first edition

\bibitem[\protect\astroncite{{Stetson}}{1987}]{Stetson:1987}
{Stetson}, P.~B.,  1987, \pasp, 99, 191

\bibitem[\protect\astroncite{{van Dyk} et~al.}{2001}]{2001ApJ...548..224V}
{van Dyk}, D.~A., {Connors}, A., {Kashyap}, V.~L., \& {Siemiginowska}, A.,
  2001, \apj, 548, 224

\bibitem[\protect\astroncite{{Weisskopf} et~al.}{2007}]{Weisskopf:2007}
{Weisskopf}, M.~C., {Wu}, K., {Trimble}, V., {O'Dell}, S.~L., {Elsner}, R.~F.,
  {Zavlin}, V.~E., \& {Kouveliotou}, C.,  2007, \apj, 657, 1026

\end{thebibliography}
\clearpage{}
\begin{table}[h!]
\centering \caption{\label{tab1} Symbols and Definitions}

\begin{tabular}{|l|p{0.75\textwidth}|}
\hline 
Symbol & Definition\tabularnewline
\hline 
$x,y$  & Image Pixel Coordinates\tabularnewline
$X_{i},Y_{i}$  & True Source Position for source $i$ on the image\tabularnewline
$psf(X_{i},Y_{i},x,y)dx\, dy$  & Telescope Point Spread Function, i.e., the probability that a photon
from a source at location $X_{i},Y_{i}$ will be detected within area
$dx\, dy$ at location $x,y$\tabularnewline
$R_{s_{i}}$  & Source Aperture for source $i$\tabularnewline
$R_{b}$  & Compound Background Aperture, common to all sources\tabularnewline
$\Omega_{s_{i}}$  & Area of Source Aperture for source $i$ (e.g. $pixel~^{2}$)\tabularnewline
$\Omega_{b}$  & Area of Background Aperture\tabularnewline
$C_{i}$  & Total Counts in Source Aperture $i$\tabularnewline
$B$  & Total Counts in Background Aperture\tabularnewline
$s_{i}$  & Net Source Counts for source $i$\tabularnewline
$b$  & Background Density (e.g. $counts-pixel^{-2}$)\tabularnewline
$f_{ij}$  & Fraction of PSF for source $j$ enclosed in source aperture $R_{s_{i}}$,
e.g., $\int_{R_{s_{i}}}psf(X_{j},Y_{j},x,y)dx\, dy$\tabularnewline
$g_{i}$  & Fraction of PSF for source $i$ enclosed in $R_{b}$, e.g., $\int_{R_{b}}psf(X_{i},Y_{i},x,y)dx\, dy$\tabularnewline
$Pois(n|\mu)$  & Probability of obtaining n counts from a Poisson Distribution with
mean $\mu$, $Pois(n|\mu)=\mu^{n}e^{-\mu}/n!=\mu^{n}e^{-\mu}/\Gamma(n+1)$\tabularnewline
$\mu_{s_{i}}$  & Expected total counts in Source Aperture $i$\tabularnewline
 & $\mu_{s_{i}}~=~\sum_{j=1}^{n}~f_{ij}s_{j}~+~\Omega_{s_{i}}b$\tabularnewline
$\mu_{b}$  & Expected total counts in Background Aperture \tabularnewline
 & $\mu_{b}~=~\sum_{i=1}^{n}~g_{i}s_{i}~+~\Omega_{b}b$\tabularnewline
\hline 
\end{tabular}
\end{table}
\clearpage{}
\begin{sidewaystable}
\centering

\caption{\label{tab:region-data}Aperture Data for Sources in Figure \ref{fig:four_source_fig}}

\begin{tabular}{|c|c|c|c|c|c|c|c|}
\hline 
\multirow{2}{*}{{\footnotesize{CSC Source CXO}}} & \multirow{2}{*}{{\footnotesize{Region ID}}} & \multicolumn{4}{c|}{{\footnotesize{PSF Contribution from Source}}} & \multirow{2}{*}{{\footnotesize{Area ($pix^{2}$)}}} & \multirow{2}{*}{{\footnotesize{Counts}}}\tabularnewline
\cline{3-6} 
 &  & {\footnotesize{J004248.4+412521}} & {\footnotesize{J004255.3+412556}} & {\footnotesize{J004251.7+412633}} & {\footnotesize{J004253.6+412550}} &  & \tabularnewline
\hline 
{\footnotesize{J004248.4+412521}} & {\footnotesize{r0115}} & {\footnotesize{0.98}} & {\footnotesize{0.00}} & {\footnotesize{0.00020}} & {\footnotesize{0.00058}} & {\footnotesize{2912.72}} & {\footnotesize{2395}}\tabularnewline
\hline 
{\footnotesize{J004255.3+412556}} & {\footnotesize{r0116}} & {\footnotesize{0.00}} & {\footnotesize{0.88}} & {\footnotesize{0.00078}} & {\footnotesize{0.0014}} & {\footnotesize{3551.00}} & {\footnotesize{759}}\tabularnewline
\hline 
{\footnotesize{J004251.7+412633}} & {\footnotesize{r0123}} & {\footnotesize{0.00}} & {\footnotesize{0.00039}} & {\footnotesize{0.96}} & {\footnotesize{0.00097}} & {\footnotesize{3120.61}} & {\footnotesize{90}}\tabularnewline
\hline 
{\footnotesize{J004253.6+412550}} & {\footnotesize{r0150}} & {\footnotesize{0.00019}} & {\footnotesize{0.098}} & {\footnotesize{0.00059}} & {\footnotesize{0.97}} & {\footnotesize{3959.92}} & {\footnotesize{273}}\tabularnewline
\hline 
{\footnotesize{--}} & {\footnotesize{Background}} & {\footnotesize{0.0072}} & {\footnotesize{0.013}} & {\footnotesize{0.029}} & {\footnotesize{0.013}} & {\footnotesize{131014.00}} & {\footnotesize{1043}}\tabularnewline
\hline 
\end{tabular}
\end{sidewaystable}
\clearpage{}

\clearpage{}
\clearpage{}
\clearpage{}
\begin{figure}
\centering

\includegraphics{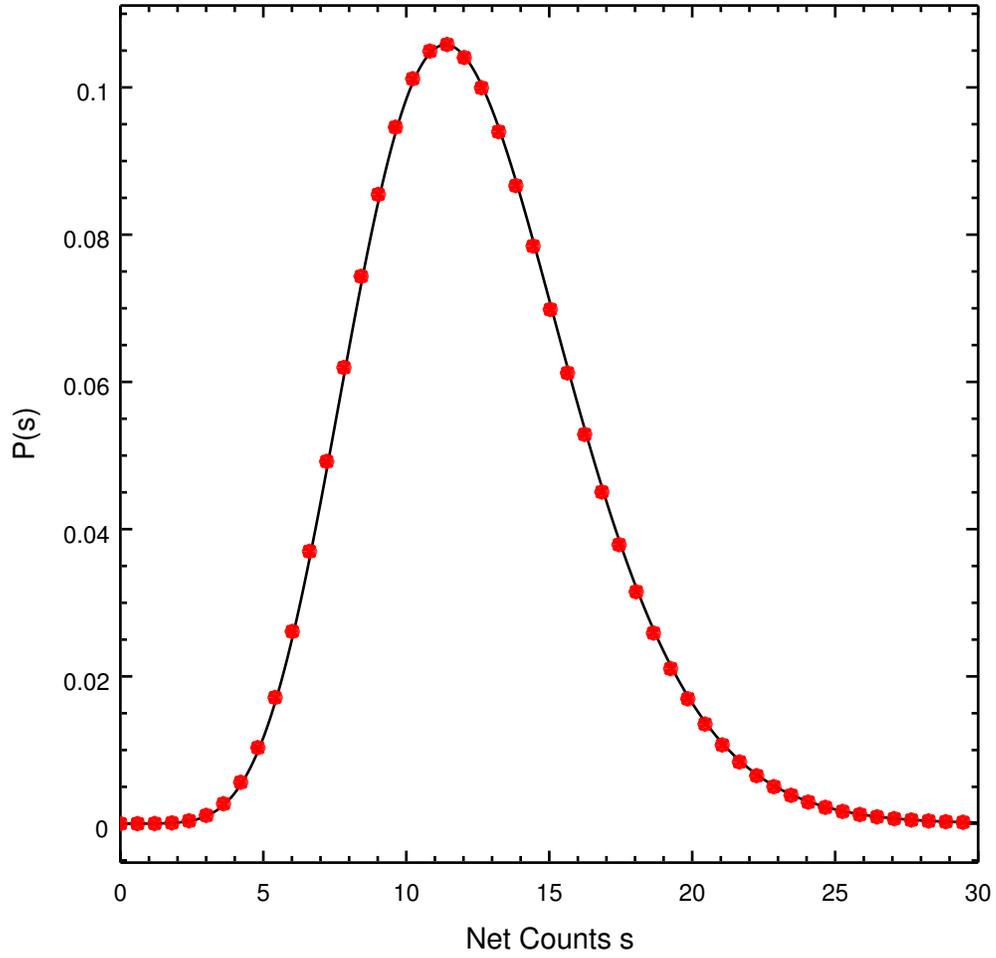}

\caption{\textit{Comparison of posterior distributions computed from Equations
\ref{eq:15} (solid black line) and \ref{eq:4} (red circles) for
the example shown in Figure \ref{fig1}. }}
\end{figure}
\clearpage{}

\begin{figure}
\centering

\includegraphics{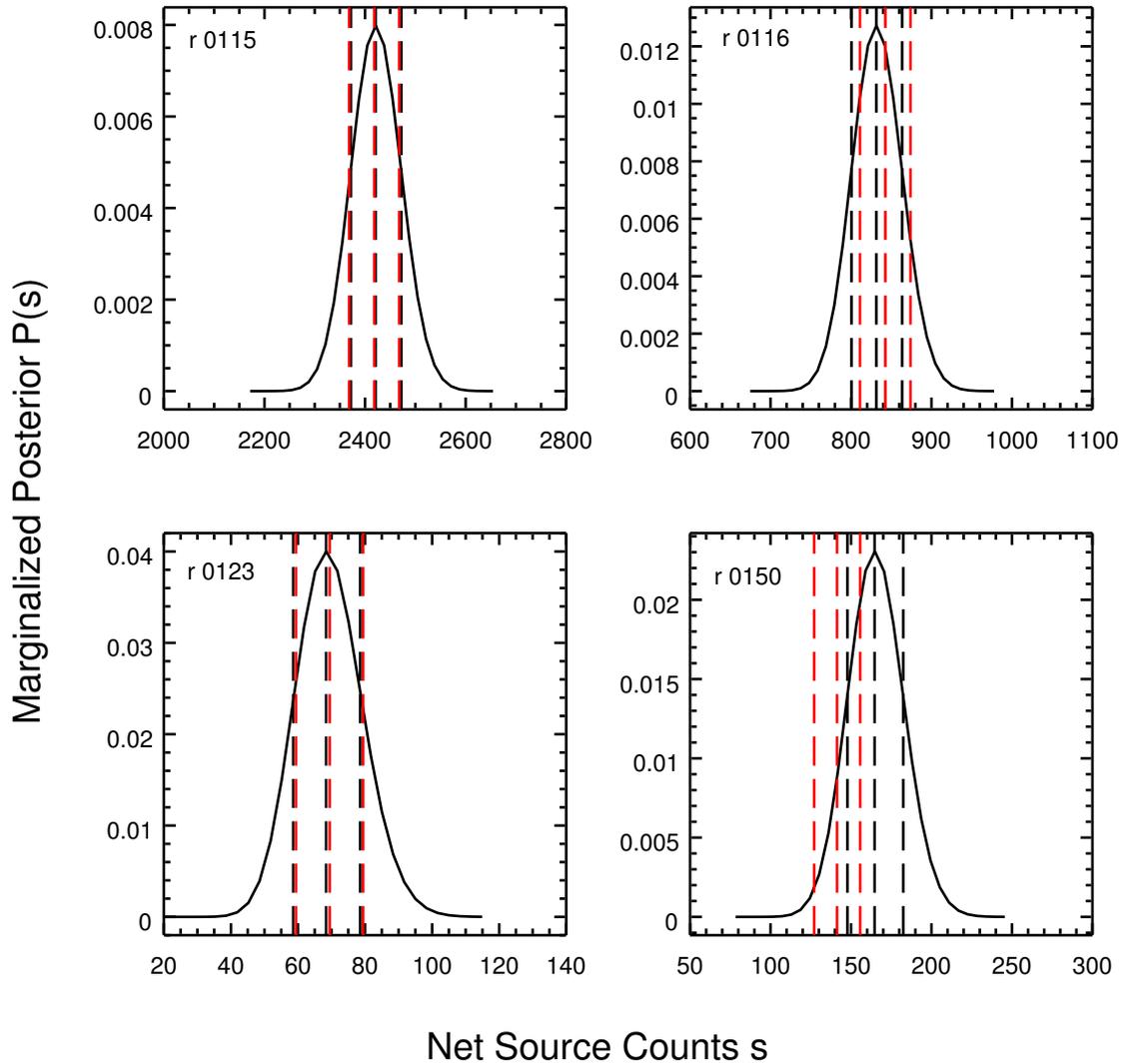}

\caption{\textit{Posterior distributions for the} \textit{four sources in Figure
\ref{fig:four_source_fig}. Modes and 68\% confidence bounds are indicated
by black vertical dashed lines. Results from Release 1.1 of the Chandra
Source Catalog are shown in red.\label{fig:posterior-four}}}
\end{figure}
\clearpage{}

\begin{figure}
\centering

\includegraphics{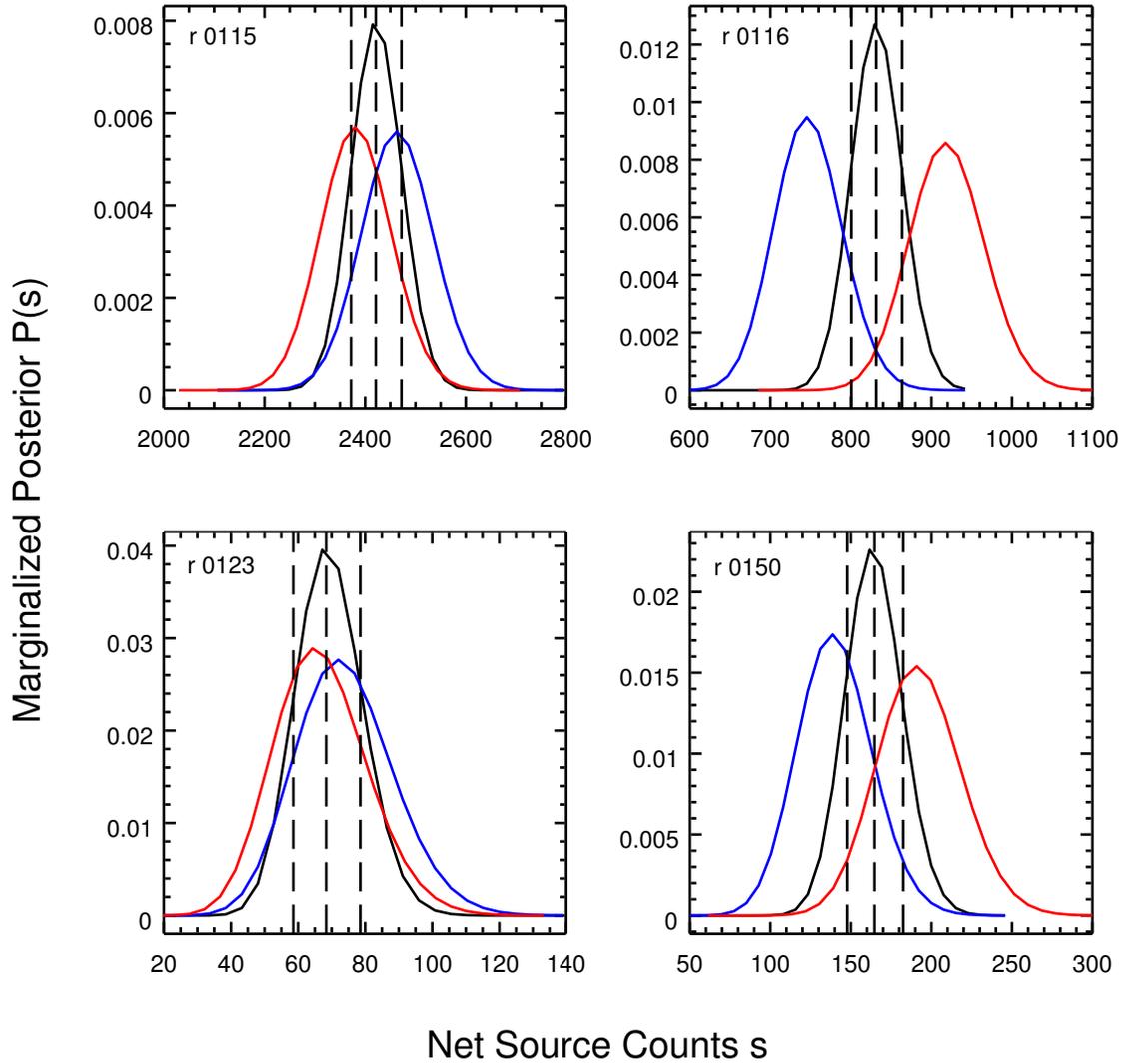}

\caption{\textit{Posterior distributions for the} \textit{four sources in Figure
\ref{fig:four_source_fig}, computed using informative priors. The red and blue
curves are for the first and second halves of the data set, computed using
non-informative priors, and the black curves are the posterior distributions for
the second half, using informative priors derived from the red
curves. Vertical dashed lines indicate the mode and 68\% confidence bounds
computed from the entire data set, using non-informative priors (see Figure~\ref{fig:posterior-four}).
\label{fig:inform-priors}}}
\end{figure}
\clearpage{}

\begin{figure}
\centering

\includegraphics[scale=0.9]{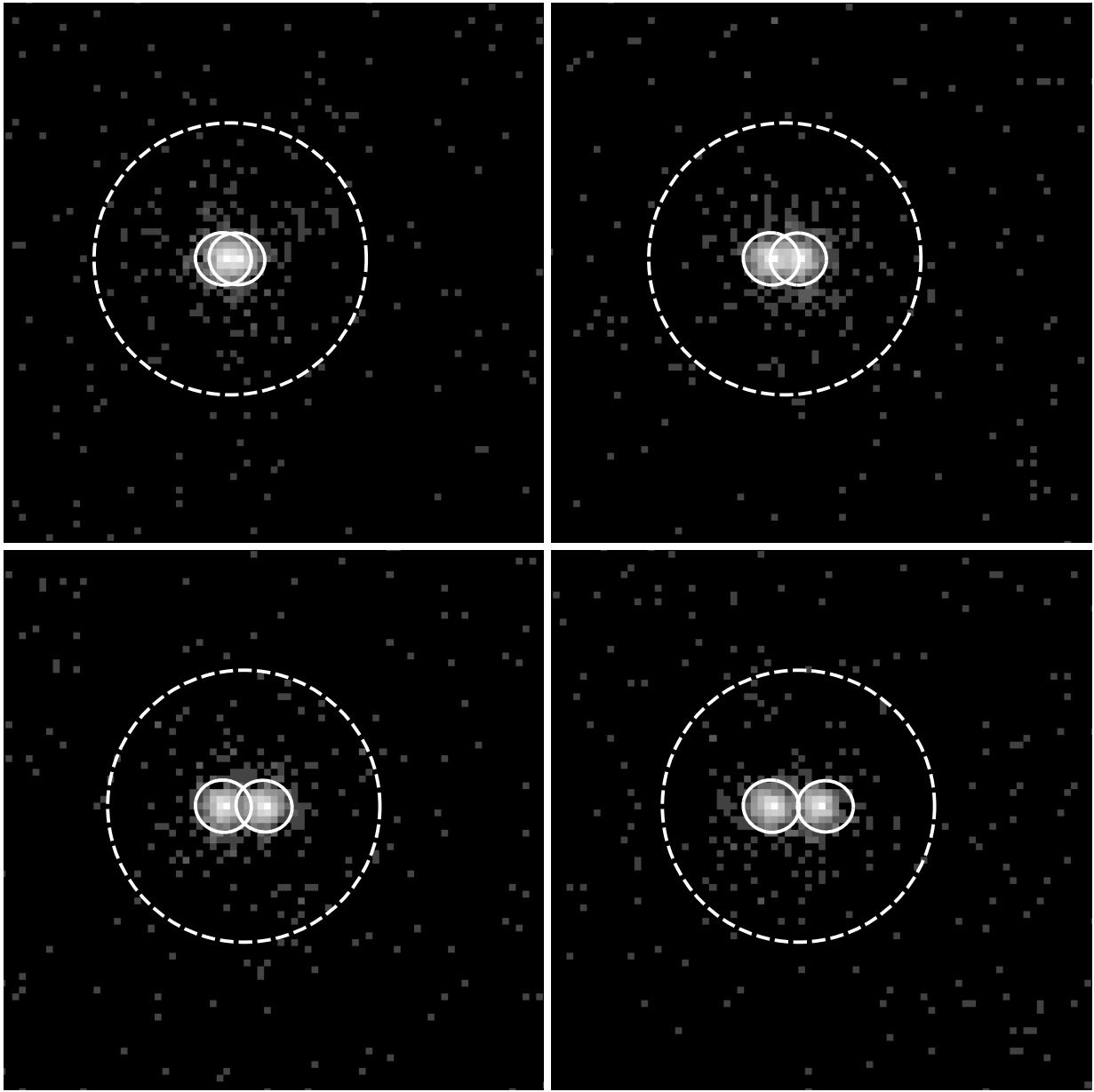}

\caption{\textit{Simulated Chandra images of two point sources separated by
}$\Delta=0.5,1.0,1.5,2.0\times r_{90}$\textit{ at an off-axis angle
of $\sim0.5^{\prime}$. Each source has a true intensity of 1000 counts,
and the mean background in the source }\emph{aperture}\textit{ is
$\sim1$ count. Source }\emph{aperture}\textit{s are constructed to
enclose approximately 90\% of the point spread function, and the background
}\emph{aperture}\textit{ (dashed circle with source }aperture\textit{s
excluded) has an area 25 times greater than that of a single source
}\emph{aperture}\textit{ and is centered at a position halfway between
the sources. \label{fig:sim04}}}
\end{figure}
\clearpage{}

\begin{sidewaysfigure}
\centering

\includegraphics[width=0.45\textheight]{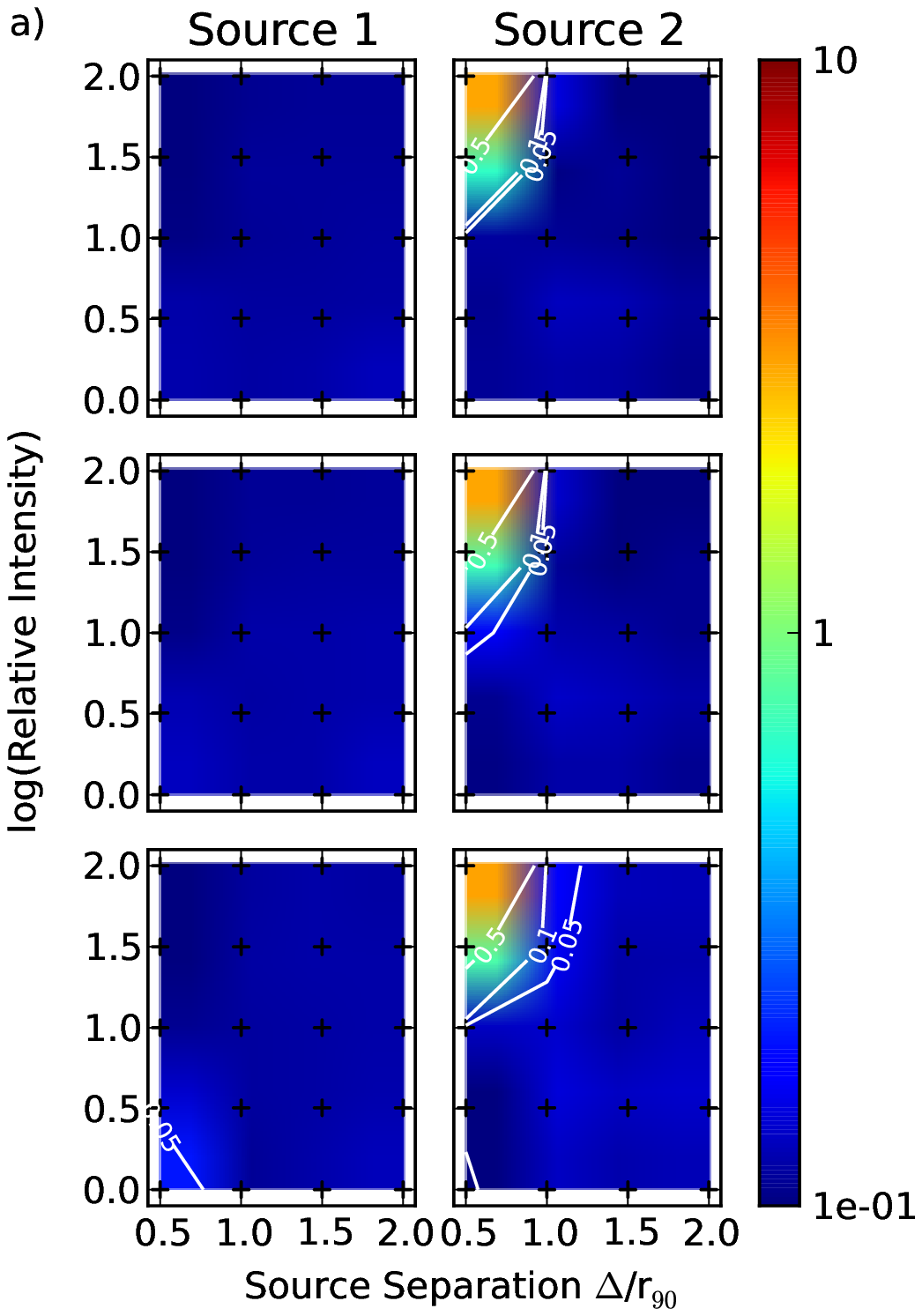}
\includegraphics[width=0.45\textheight]{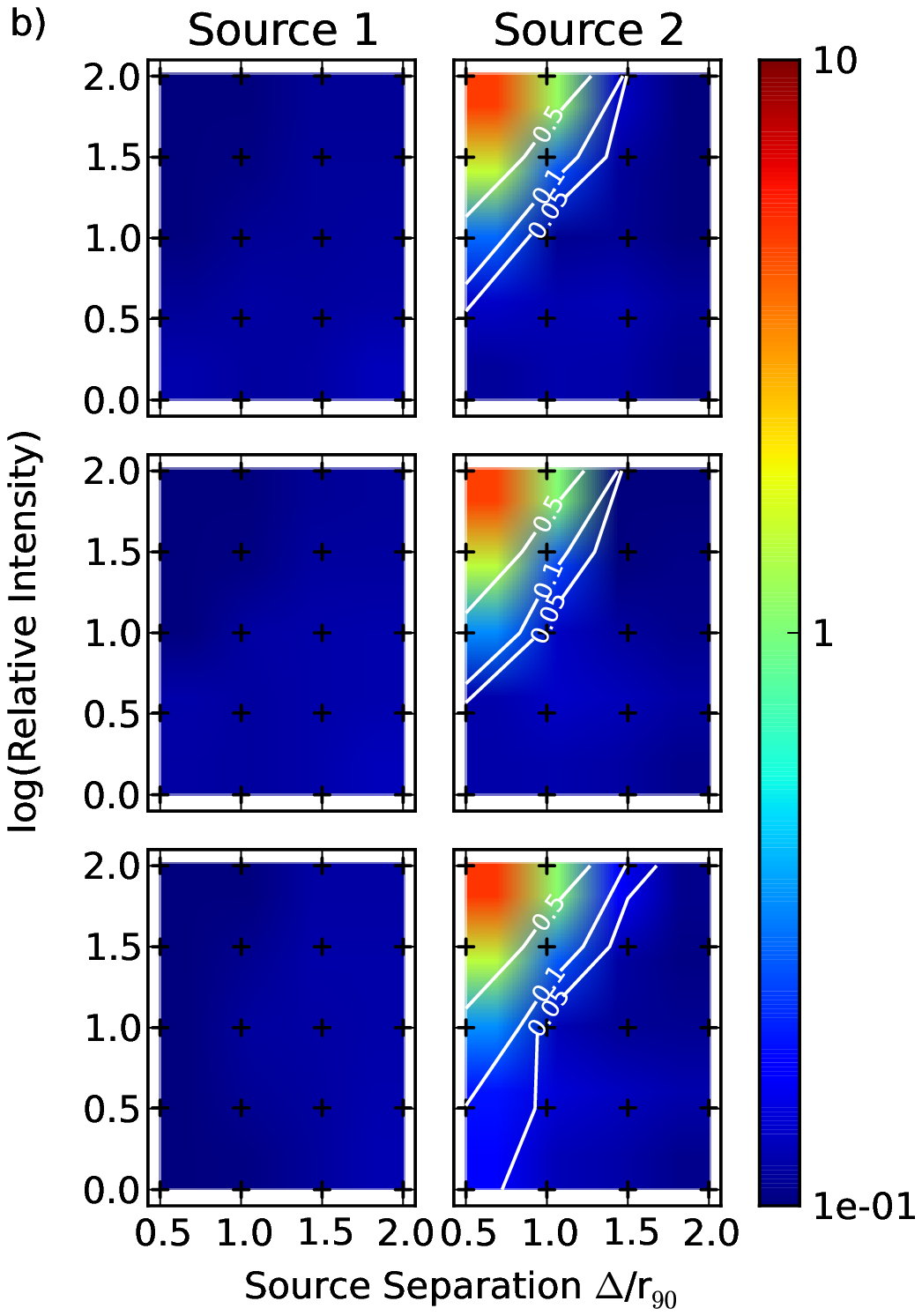}

\caption{\emph{Average fractional error in source intensity as a function of
$log_{10}r$ and $\Delta$ for relative background
$b$ of $0.001,\,0.01,\,$ and $0.100$, from top to bottom. Contours for fractional errors of $-0.05,\,0.05,\,0.1,\,0.5,$
and $5.0$ are indicated. Sampled values are indicated by crosses and the
interpolated surface is displayed using a logarithmic colormap. a) Case 1: overlap
area in source apertures is assigned to the
aperture for Source 1. b) Case 2: overlap area is assigned to the aperture for Source 2.\label{fig:S1full_frac_err}}}
\end{sidewaysfigure}
\clearpage{}

\begin{sidewaysfigure}
\centering

\includegraphics[width=0.45\textheight]{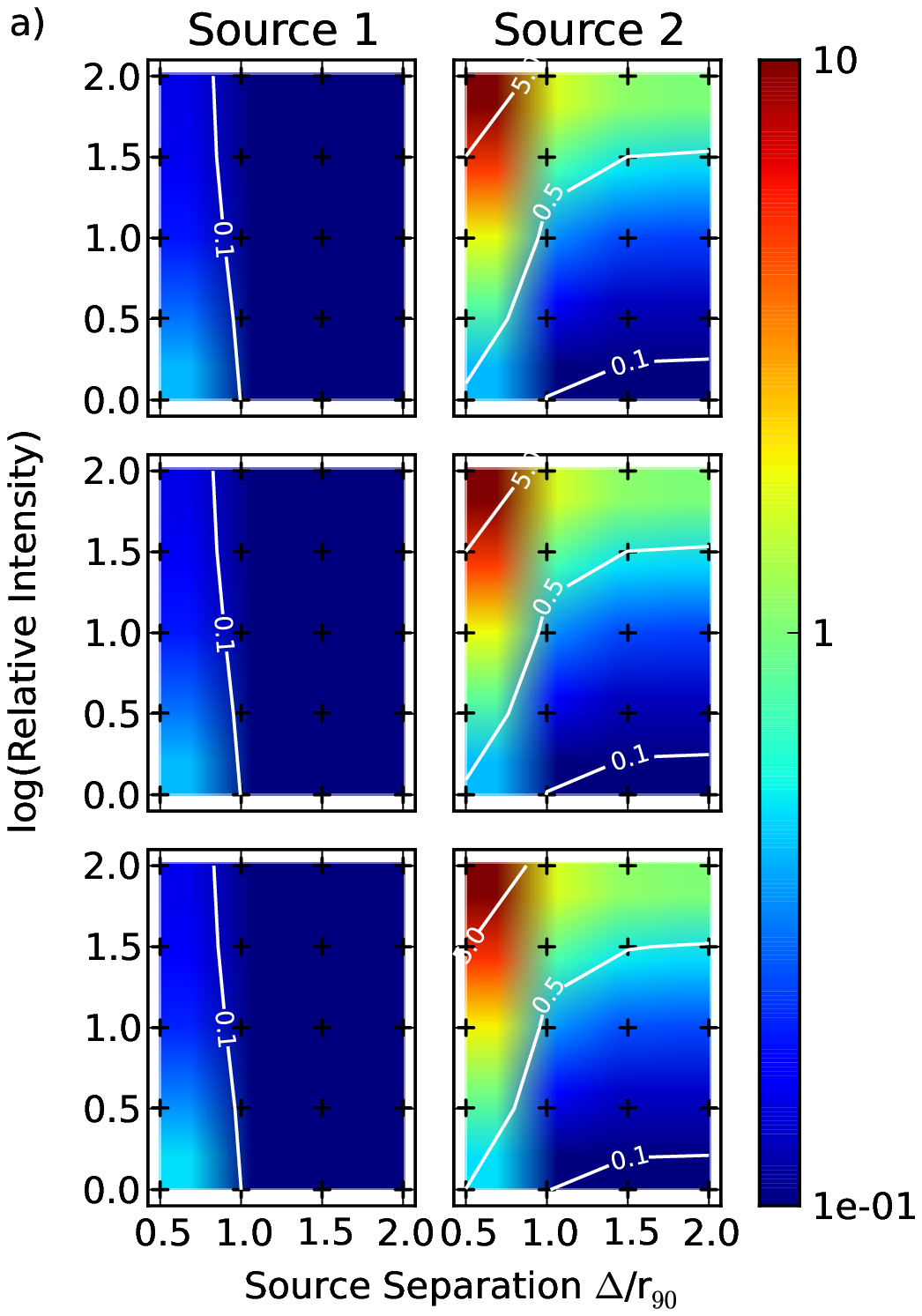}
\includegraphics[width=0.45\textheight]{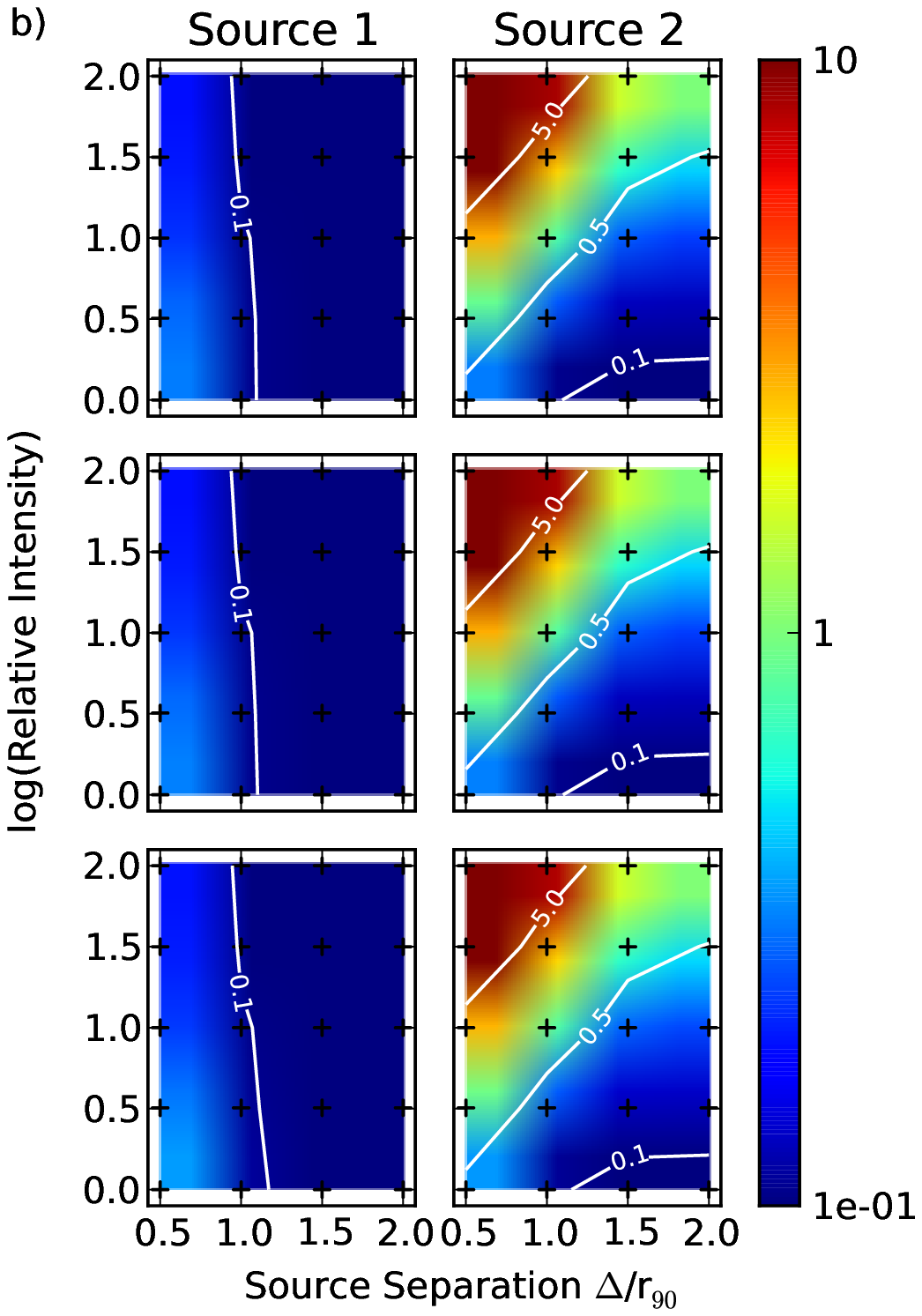}

\caption{\emph{Average fractional width of source intensity probability
    distributions (see Figure~\ref{fig:S1full_frac_err} for plot details). a) Case 1: overlap
area in source apertures is assigned to the
aperture for Source 1. b) Case 2: overlap area is assigned to the aperture for Source 2.
\label{fig:Average-fractional-width}}}
\end{sidewaysfigure}
\clearpage{}
\clearpage{}
\begin{sidewaysfigure}
\centering

\includegraphics[width=0.45\textheight]{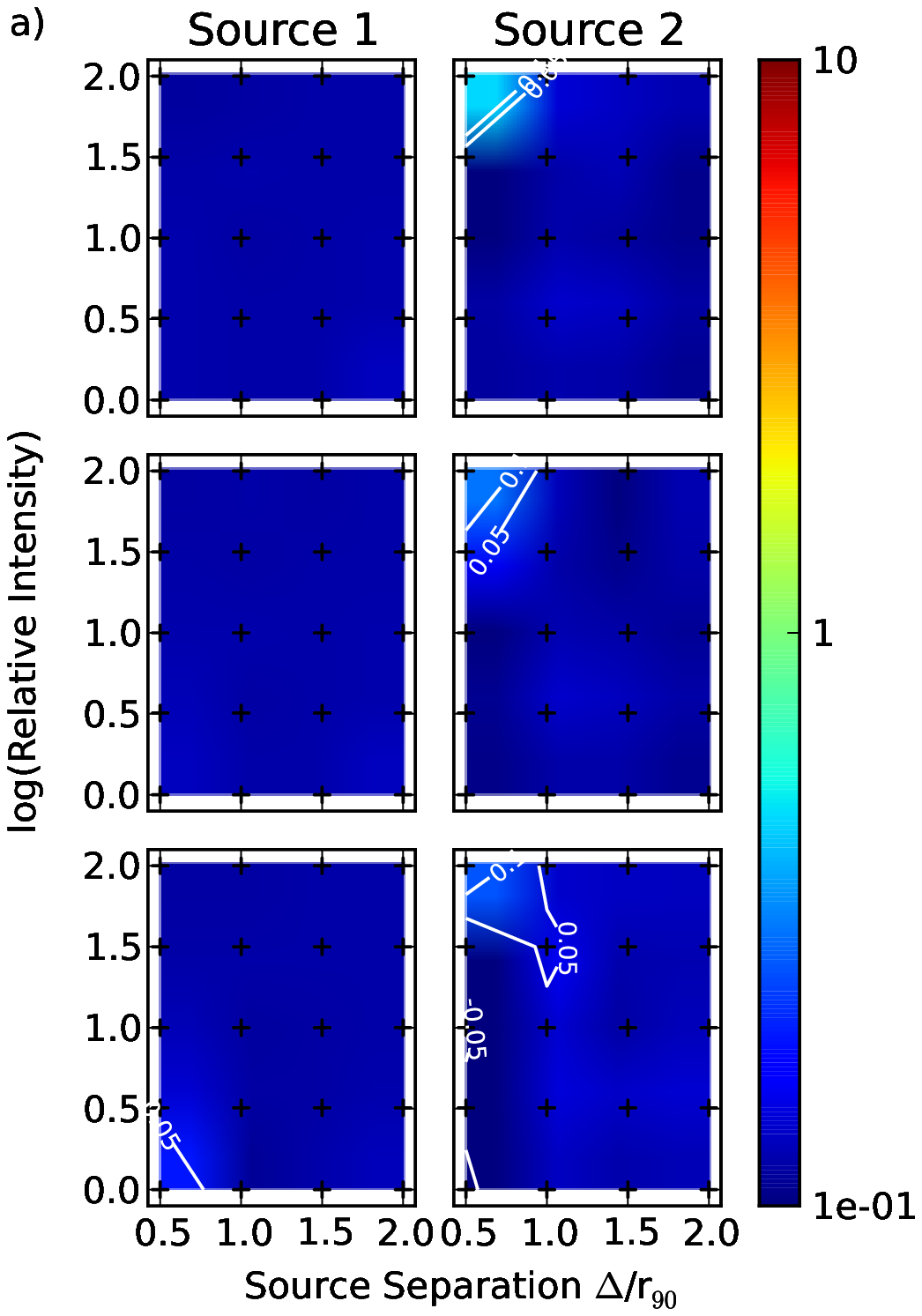}
\includegraphics[width=0.45\textheight]{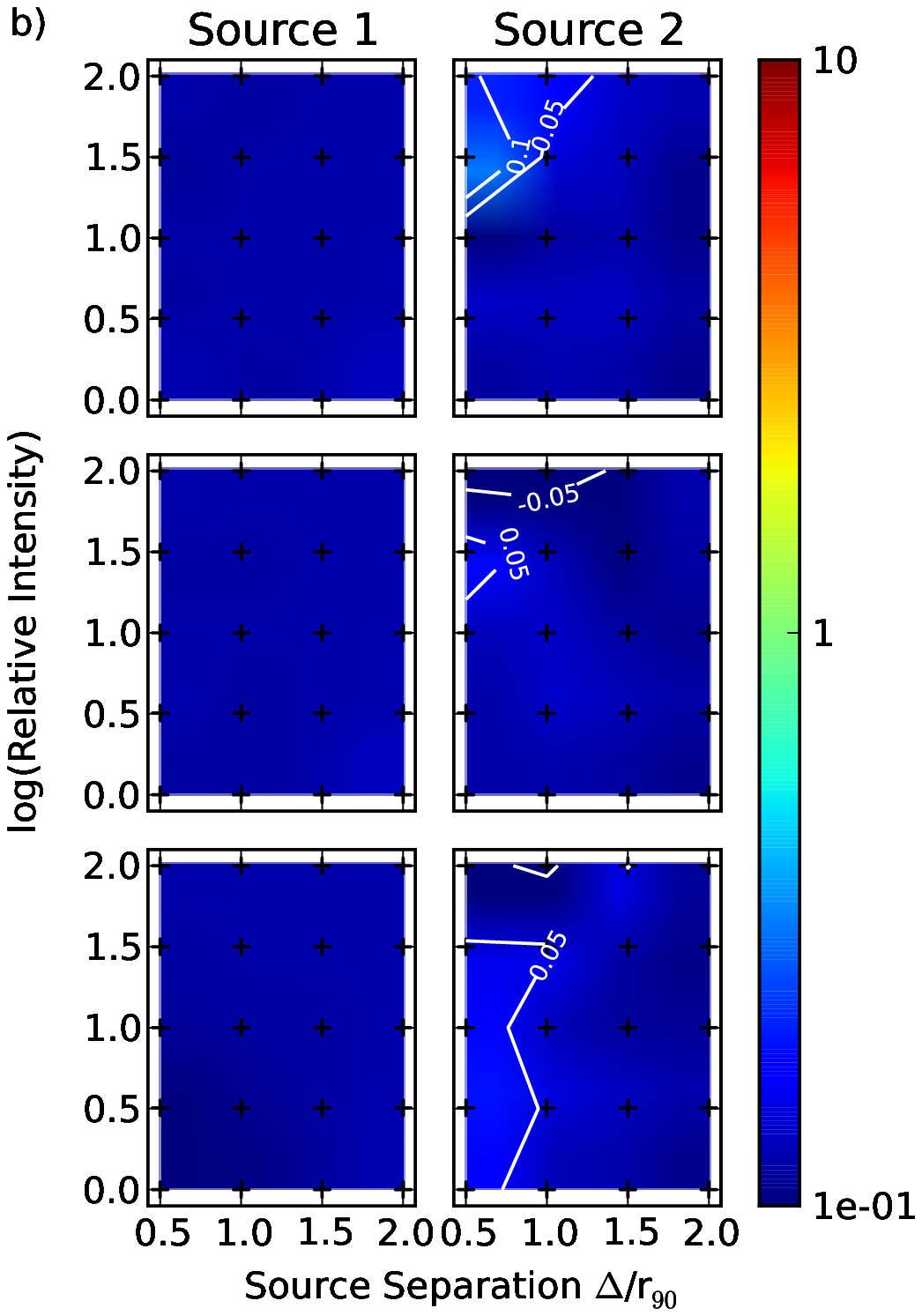}

\caption{\emph{Average fractional errors, as in Figure~\ref{fig:S1full_frac_err}
based on maximum-likelihood determinations for source intensities and errors (see Figure~\ref{fig:S1full_frac_err} for plot details).
 a) Case 1: overlap
area in source apertures is assigned to the
aperture for Source 1. b) Case 2: overlap area is assigned to the aperture for Source 2.\label{fig:mle-frac-err}}}
\end{sidewaysfigure}
\clearpage{}
\clearpage{}

\begin{sidewaysfigure}
\centering

\includegraphics[width=0.45\textheight]{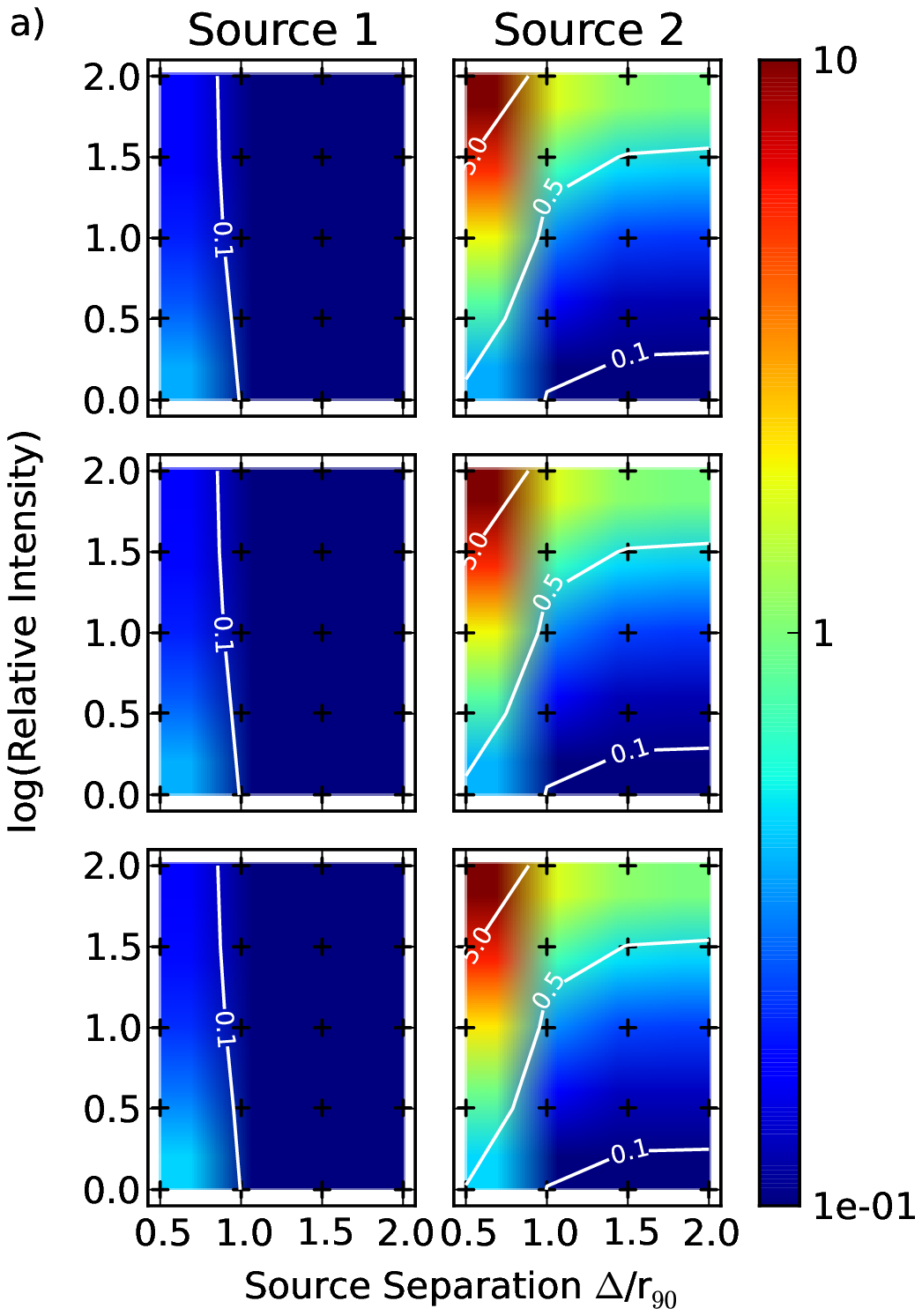}
\includegraphics[width=0.45\textheight]{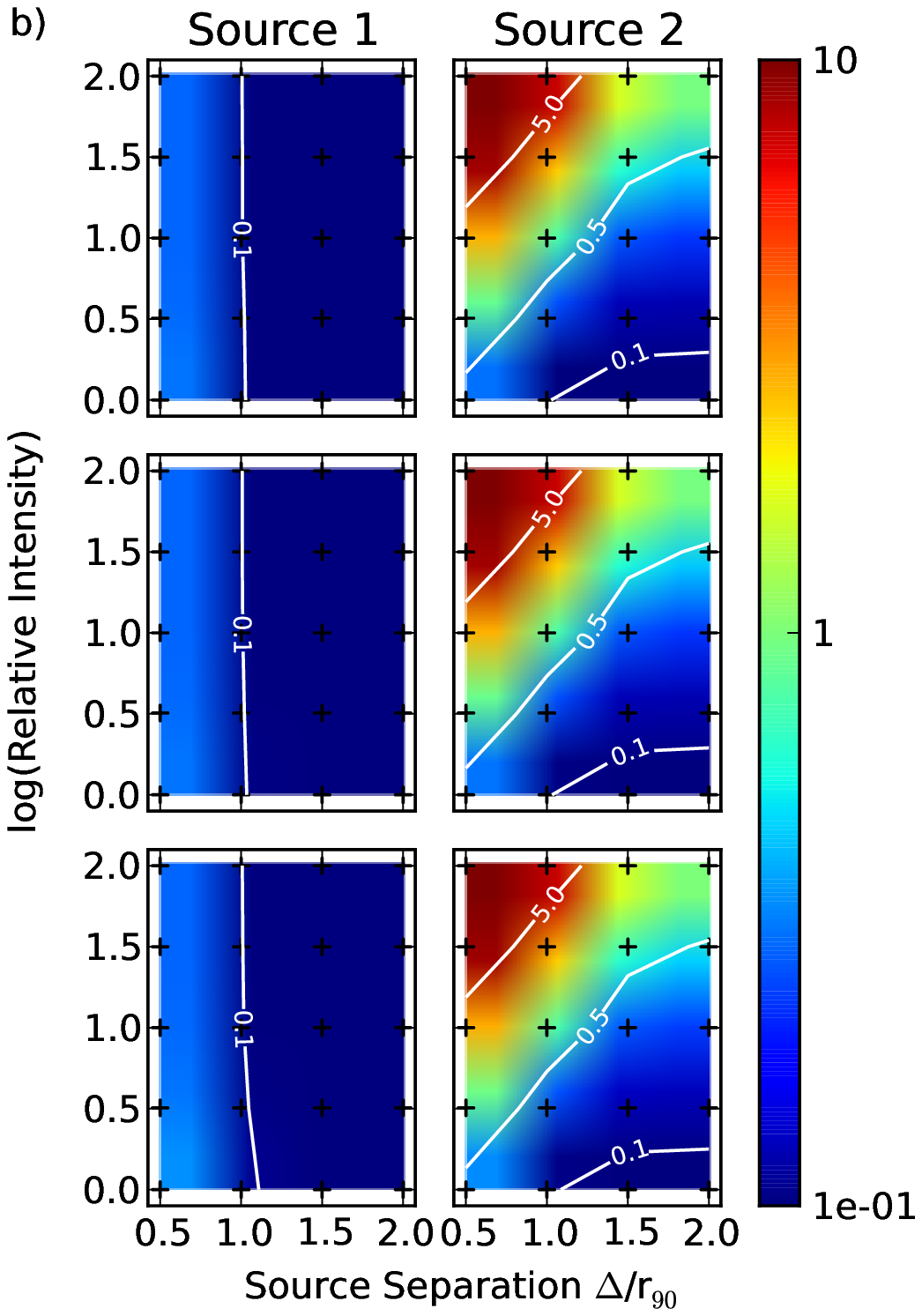}

\caption{\emph{Average fractional width of source intensity probability distributions,
based on maximum-likelihood determinations for source intensities and errors (see Figure~\ref{fig:S1full_frac_err} for plot details).
a) Case 1: overlap
area in source apertures is assigned to the
aperture for Source 1. b) Case 2: overlap area is assigned to the aperture for Source 2.\label{fig:mle-frac-width}}}
\end{sidewaysfigure}
\clearpage{}

\begin{sidewaysfigure}
\centering

\includegraphics[width=0.45\textheight]{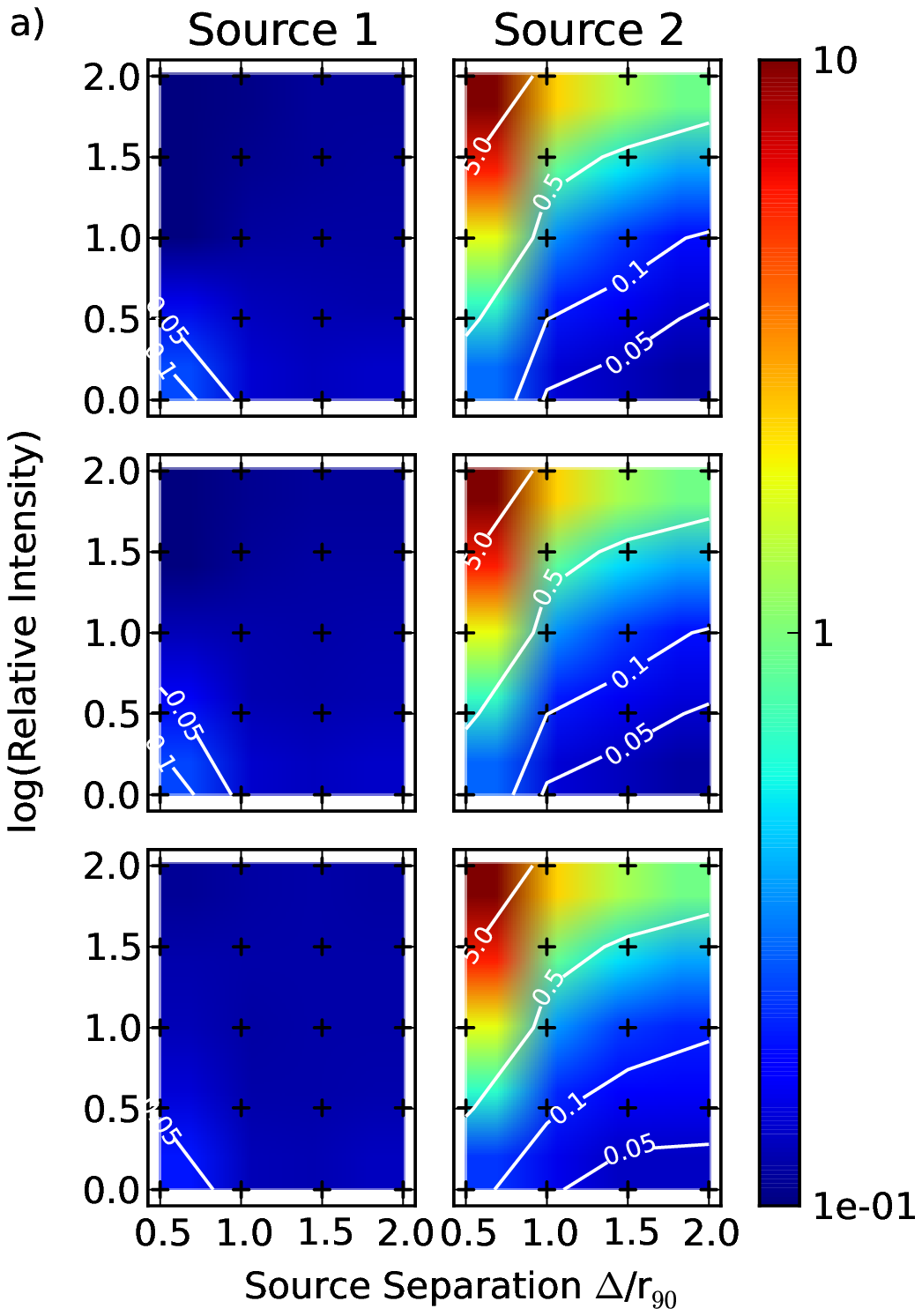}
\includegraphics[width=0.45\textheight]{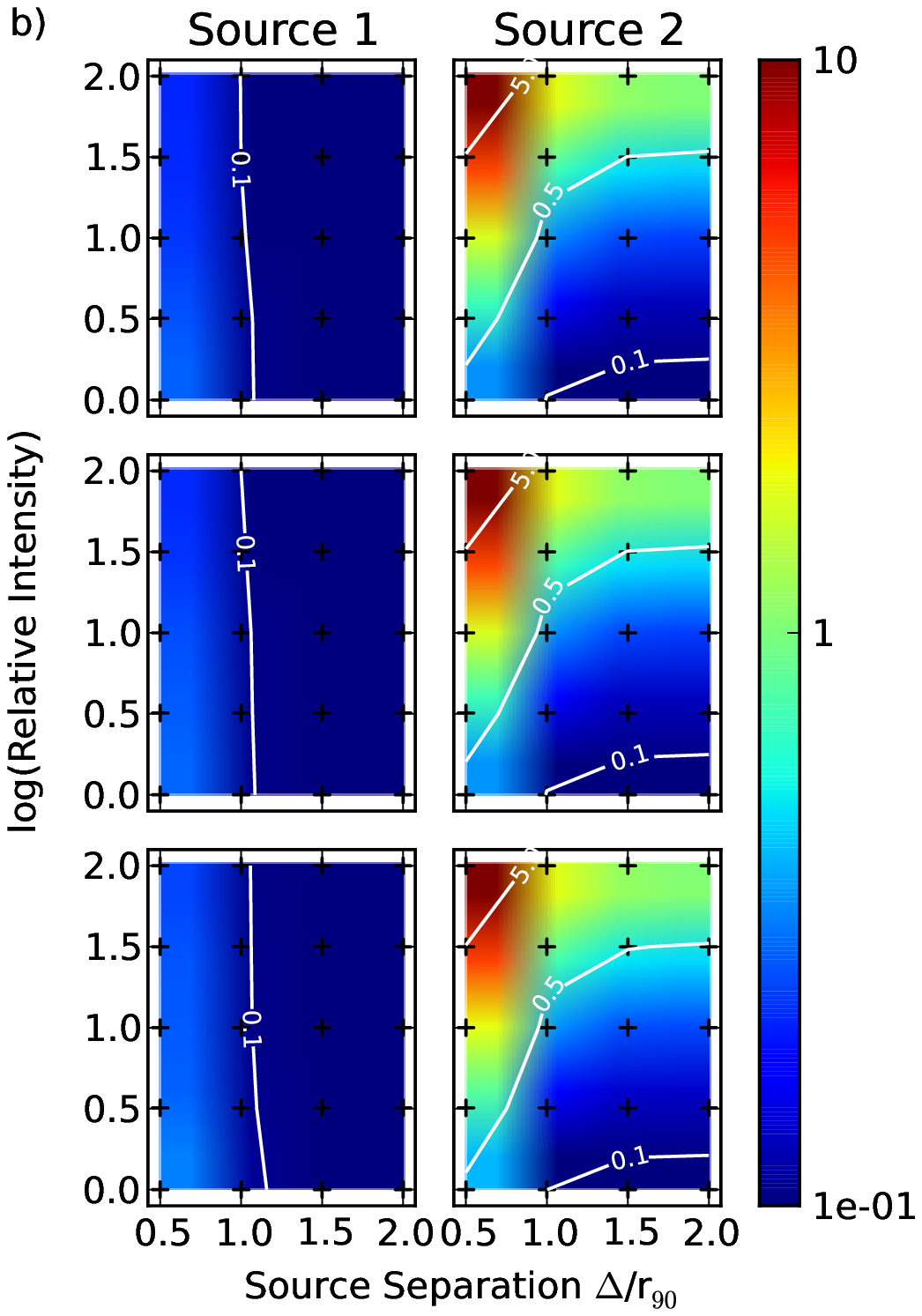}

\caption{\emph{Average fractional errors and widths of source intensity probability distributions,
assuming source apertures used in Rel. 1.1 of the Chandra Source Catalog.
a) Average fractional errors. b) Average fractional widths.\label{fig:aprates}}}
\end{sidewaysfigure}











\end{document}